\documentclass[showpacs,aps,prd,floatfix,amsmath,amssymb]{revtex4}
\usepackage{graphicx}
\usepackage{amsmath}
\usepackage{color}
\usepackage[utf8x]{inputenc}

\begin{document}

\title{Chromomagnetic and chromoelectric dipole moments of the top quark in
    the fourth-generation THDM}
\author{A. I. Hern\'andez-Ju\'arez}
\affiliation{Facultad de Ciencias F\'isico Matem\'aticas, Benem\'erita Universidad Aut\'onoma de Puebla, Apartado Postal 1152, Puebla, Puebla, M\'exico}
\author{A. Moyotl}
\affiliation{Ingenier\'ia en Mecatr\'onica, Universidad Polit\'ecnica de Puebla, Tercer Carril del Ejido Serrano s/n, San Mateo Cuanal\'a, Juan C. Bonilla,  CP 72640, Puebla, Puebla, M\'exico}
\author{G. Tavares-Velasco}
\affiliation{Facultad de Ciencias F\'isico Matem\'aticas, Benem\'erita Universidad Aut\'onoma de Puebla, Apartado Postal 1152, Puebla, Puebla, M\'exico}
\date{\today}

\begin{abstract}
The chromomagnetic dipole moment (CMDM) and chromoelectric dipole moment (CEDM) of  the top quark are calculated at the one-loop level in the framework of the two-Higgs doublet model with   four fermion generations (4GTHDM), which is still consistent with experimental data and apart from  new scalar bosons ($H^0$, $A^0$, and $H^\pm$) and quarks  ($b'$ and $t'$) predicts  new sources of $CP$ violation via  the extended $4\times 4$ CKM matrix.  Analytical expressions for  the CMDM and CEDM of a quark  are presented both in terms of Feynman parameter integrals, which are explicitly integrated,  and Passarino-Veltman scalar functions, with the main contributions arising from loops carrying the scalar bosons accompanied by the third- and fourth-generation quarks. The current bounds on the parameter space of the 4GTHDM are discussed and a region  still consistent with the LHC data on the 125 GeV Higgs boson and the oblique parameters is identified. It is found that  the top quark CMDM, which is induced by all the scalar bosons, can reach values of the order of $10^{-2}$--$10^{-1}$. As for the   top quark CEDM, it only receives contributions from the charged scalar boson and can reach values of the order of $10^{-20}$--$10^{-19}$  ecm for relatively light $m_{H^\pm}$ and heavy $m_{b'}$, with the dominant contribution arising from the $b$ quark. The CEDM would be  the most interesting prediction of this model as it can be larger than the value predicted by the usual THDMs by one order of magnitude.
\end{abstract}


\date{\today}

\maketitle

\section{Introduction}
\label{Intro}

Since its discovery in 1995 by the CDF and D0 experiments at Fermilab's Tevatron \cite{Abe:1995hr,D0:1995jca}, the top quark has played a special role in the study of the phenomenology  of the standard model (SM), which stems from the fact that its  mass is of the order of  the electroweak symmetry breaking scale.
Even more, the top quark is unique as it does not hadronize unlike all other quarks, due to its tiny lifetime $\tau_t=5\times 10^{-25}$ s, but it also can decay semi-weakly and has a Yukawa coupling of the order of the unity.
At the CERN Large Hadron Collider (LHC),  the top quark is pair produced mainly via the processes  $q\overline{q}\rightarrow t\overline{t}$ and $gg\rightarrow t\overline{t}$. At  a center-of-mass energy   $\sqrt{s}=14$ TeV, about 90\% of the top quark production arises from gluon fusion and the remainder from $q\overline{q}$ annihilation \cite{HiggsProduction1}. The LHC is thus a top quark factory, which  opens up  a plethora of opportunities to test its  properties:  mass, couplings to other SM particles, spin observables, rare decays, etc. A top quark factory also provides a laboratory to search for new physics effects. Along these lines,  the study of the new contributions to the chromomagnetic dipole moment (CMDM) and chromoelectric dipole moment (CEDM) of the top quark is a topic worth studying as they could be at the reach of experimental measurement in the near future.

In the context of the SM, there are many unsolved problems.  Among them, one of the most interesting is the baryon asymmetry of the universe. According to Sakharov's criteria \cite{BaryonAsy}, $CP$ violation is a necessary requirement for this phenomenon. In the SM, the complex phase of the Cabibbo-Kibayashi-Maskawa (CKM) matrix \cite{CKM1, CKM2} gives rise to $CP$ violation, though it is still  not enough to explain the baryon asymmetry, which means that new sources of $CP$ violation  beyond the SM are required. It is therefore necessary to search for evidences of any $CP$-violating effects. We are thus interested in looking for evidences of such effects through the  $t\overline{t}g$ vertex, whose anomalous contributions can be written via the following dimension-five effective Lagrangian
\begin{equation}
\mathcal{L}=-\frac{ g_s T^{a}} {2}\bar{t}\,\frac{a_t}{2m_t} \sigma^{\mu\nu}\, tG^a_{\mu\nu}-\frac{ T^{a}} {2}\bar{t}\,i\sigma^{\mu\nu}   \gamma ^5 d_t \,tG^a_{\mu\nu},
\end{equation}
where $G^a_{\mu\nu}$ is the gluon  strength tensor and $T^a$ are the $SU(3)$ generators. The anomalous couplings $a_t$ and $d_t$ are known as the CMDM and CEDM, respectively, though alternative definitions for the latter are also used in the literature \cite{Bernreuther:2013aga}. The existence of a  CEDM   implies time-reversal violation, which is equivalent to $CP$ violation because of the $CPT$ theorem, so any evidence of a CEDM of the top quark would indicate a  $CP$-violating effect.   In the SM, the top quark CMDM is induced at the one-loop level { and its value at the leading order is $-5.6\times 10^{-2}$, with the electroweak (EW) and quantum chromodynamics (QCD) contributions being  $-6.4\times 10^{-2}$ and $7.5 \times 10^{-3}$, respectively \cite{ManyModels}} . {As for the CEDM, it arises at three loops  \cite{3loop} and  its value has been estimated to be negligibly small, of the order of $10^{-30}\; {\rm g}_s$cm \cite{Soni:1992tn}} , therefore a sizeable CEDM would hint  new sources of $CP$ violation.

{ Constraints on the top CMDM and CEDM has been set \cite{Constricciones} using the ATLAS data on the $\bar{t}t $ production cross section through the lepton plus jet channel.  The corresponding bounds  are $-0.034<a_t<0.031$ and $\left|d_t\right| <2.17\times10^{-16}$ ecm \cite{Constricciones}. It is expected that the LHC data on $\sigma(pp\to \bar{t}t)$ at $\sqrt{s}=14$ TeV would allow to place the bounds $-0.016 \le a_t\le 0.008$  and $|d_t|\le 3.6 \times 10^{-17}$ ecm. Even more, a sensitivity to the CEDM  of the top quark of about $|d_t|\le 1.6 \times 10^{-18}$ ecm  would be reached through the measurement of a T-odd correlation in the process $pp\to \bar{t}t $ with 10 fb$^{-1}$ \cite{Constricciones}.}
It is worth contrasting these values with the electromagnetic properties of the top quark, namely, the anomalous magnetic dipole moment (MDM) and the electric dipole moment (EDM), which have also been calculated in the literature in the framework of the SM and several of its extensions.  In the SM, there are three types of contributions to the top quark MDM, namely, QED, EW, and QCD contributions, with the  total SM contribution being $3.5\times10^{-2}$ \cite{Labun:2012fg}. As far as the top quark EDM is concerned, it  has not been calculated yet  but an estimate of   about $10^{-30}$ ecm was obtained by scaling the value of the electron EDM \cite{Soni:1992tn,Hoogeveen:1990cb}.

Apart from  the SM  calculations \cite{SM, ManyModels}, the top quark CMDM and CEDM have been studied in several SM extensions, such as little Higgs models \cite{LittleHiggs, LittleHiggs2}, two-Higgs doublet models (THDMs) \cite{THDM,Gaitan:2015aia, THDMandMSSM, ManyModels,Iltan:2001vg}, the minimal supersymmetric standard model (MSSM) \cite{MSSM1, MSSM2, MSSM3,THDMandMSSM, MSSM4},  unparticles \cite{Unparticle}, technicolor   \cite{ManyModels}, 331 models \cite{ManyModels}, and models with vectorlike multiplets \cite{VecMul}.  Furthermore, phenomenological analysis of the $t\bar{t}g$ anomalous couplings has been performed in the context of single top production \cite{Singletop1, Singletop2, Singletop3, Singletop4}, top pair production \cite{Pairtop1,Pairtop2, Pairtop3, Pairtop4, Pairtop5, Pairtop6, Pairtop7, Pairtop8, Pairtop9, Pairtop10, Pairtop11, Pairtop12}, top pair plus jet production \cite{Plusjet}, direct photon production \cite{DirectPhotonPro}, spin correlation in  top pair production \cite{SpinCorrelation}, $CP$ violation in top pair production \cite{PairCPviolation}, etc.

In this work, we study the one-loop contributions to the CMDM and CEDM of the top quark in the  THDM with a   fourth family of fermions (4GTHDM), which was proposed by Bar-Shalom {\it et al.} in 2011 \cite{4G2HDM}. A fourth SM-like fermion family was introduced in the past in the so-called sequential SM  (SM4) \cite{SM4}, which is  the most simple extension of the SM with additional up-type and down-type quarks denoted by $t^\prime$ and $b^\prime$, respectively. The introduction of a new quark family requires a $4\times 4$  CKM  matrix, which can be parametrized by  six real parameters and three complex phases. The latter imply new sources of $CP$ violation as those required to solve  the baryon asymmetry puzzle. {
Although there is no symmetry that  prevents the SM from being extended with extra SM-like fermion families,} a fourth generation of such fermions has been ruled out  by the measurement of the invisible decay width of the $Z$ gauge boson, which is consistent with three flavors of light neutrinos \cite{Z}, though extra neutrinos with mass $m_{\nu^\prime}>m_Z/2$ are still allowed. However, the SM4 is not compatible with the LHC data on Higgs boson production \cite{NoSM4, NooSM4, NoooSM4, NooooSM4,Eberhardt:2012gv}  as an extra family of quarks with SM-like couplings would increase the Higgs production via gluon fusion \cite{SM4Higgs} at a level not consistent with that experimentally observed \cite{HiggsProduction2}. In contrast,
the 4GTHDM is still  consistent with the 125 GeV Higgs boson discovered in 2012 \cite{1254G2HDM}: the theoretical prediction for Higgs boson production at the LHC agrees with that observed in a certain region of the parameter space of the model. This was shown by the authors of the 4GTHDM in  Refs. \cite{Review4G2HDM,1254G2HDM}, where they perform a fit to the parameters of the lightest scalar boson $h^0$ with the LHC data on the 125 GeV Higgs boson to constrain the masses of the quarks of the fourth family and other parameters of the model. Along this line, other versions of THDMs with  a fourth-generation of fermions that are still compatible with LHC data have been considered in the literature \cite{He:2011ti,Das:2017mnu}.
{At the LHC, the ATLAS and CMS collaborations have searched for  new heavy quarks, but the corresponding bounds are model dependent. The current lower bound on chiral fourth-generation quarks is very stringent, namely,  $m_{t^\prime,b^\prime} \gtrsim 700$ \ GeV \cite{Aad:2015tba,Patrignani:2016xqp}, which is above the unitarity bound  $m_Q\lesssim550$ GeV \cite{Chanowitz:1978mv,Chanowitz:1978uj}. However,  the experimental constraint, obtained by assuming that the main decay channels of the heavy quarks are $b'\to Wt, bh, bZ$ and $t'\to Wb, th, tZ$, can be evaded by tuning the model parameters  \cite{Bar-Shalom:2016ehq} and thus masses within the interval of about 350--600 GeV are still allowed.}

The contributions to the MDM of a fermion  were calculated in the 4GTHDM framework prior the Higgs boson discovery \cite{Magnetic4G2HDM}, with a post-discovery update presented in  \cite{Review4G2HDM}. Furthermore, several decay modes of the top quark have been studied within this model \cite{Magnetic4G2HDM, Decay4G2HDM, Decay24G2HDM}, and  the inclusion of a fourth generation of chiral  fermions was studied in \cite{Bar-Shalom:2016ehq}. We present below an analysis of the contributions of the new heavy scalar bosons of the 4GTHDM  to the CMDM and CEDM of the top quark, along with the implications of  the presence of the quarks of a  fourth family.

The rest of this work is organized as follows. In Sec. \ref{Model} we present a brief outline of the framework of the 4GTHDM, with particular emphasis on  the Yukawa Lagrangian, from which the couplings of the new scalar bosons with the SM and fourth-generation fermions are extracted. Section \ref{CromoDipoleMoments} is devoted to the analytical results for the CMDM and CEDM of the top quark in terms of Feynman parameter integrals and Passarino-Veltman scalar functions.  In Sec, \ref{Numerical} we  discuss the most up-to-date constraints on the parameter space of the model, and  perform a numerical analysis of  the behavior of the CMDM and CEDM of  the top quark for the still allowed parameter values. The concluding remarks and outline are presented in Sec. \ref{conclusions}.

\section{The two-Higgs doublet with a  fourth generation of fermions}
\label{Model}
{The study of THDMs, which only add an extra doublet to the SM, is well motivated as they are simple but offer a great variety of new physics effects, such as new sources of $CP$ violation, new neutral and charged scalar bosons, tree-level scalar flavor-changing neutral currents (FCNCs), etc. In addition,  the MSSM scalar sector and axion models require two Higgs doublets, which have also been used to conjecture that the  top quark mass is very heavy due to a disparity between the  vacuum expectation values (VEVs) of the two Higgs doublets, namely, $ \upsilon_{h}\gg \upsilon_{\ell}$, where $\upsilon_{h}$ is the VEV of the Higgs doublet that only couples to the (heavy) top quark and $\upsilon_{\ell}$ is that of the Higgs doublet that  only couples to the remaining (light) fermions \cite{Das:1995df}.

A variant  of the usual THDMs is the so-called 4GTHDM, obtained by adding a fourth family of fermions, for which we present a short overview  and refer the interested   reader to the original References \cite{4G2HDM,1254G2HDM} for a more detailed discussion. Following  the notation of Ref. \cite{Das:1995df}, the two Higgs doublets of the 4GTHDM are denoted by
 $ \Phi_{\ell}$ and $ \Phi_{h}$, with VEVs $\upsilon_{\ell}$ and $\upsilon_{h}$, respectively. We use the  definitions $ \upsilon\equiv\sqrt{\upsilon_{\ell}^{2}+\upsilon_{h}^{2}}  $ and $\tan \beta=\upsilon_{h}/\upsilon_{\ell} $.} As already mentioned, apart from the extra Higgs doublet, in the 4GTHDM  a  fourth fermion family is introduced, which can still be in accordance with the LHC data on the 125 GeV Higgs boson \cite{1254G2HDM} and can lead to  very interesting new physics effects such as  new sources of $CP$ violation. In this model, the Yukawa Lagrangian of the quark sector can be written as follows
\begin{equation}
\label{Yukawa}
\begin{aligned}
    \mathcal{L}_Y=& \ -\overline{Q}_L\left(\Phi_\ell \mathbf{F}\cdot\left(\mathbf{I}-\mathbf{I}^{\alpha_d\beta_d}_d\right)+\Phi_h\mathbf{F}\cdot \mathbf{I}^{\alpha_d\beta_d}_d\right)d_R   \\
    &  -\overline{Q}_L\left(\tilde{\Phi}_\ell \mathbf{G}\cdot\left(\mathbf{I}-\mathbf{I}^{\alpha_u\beta_u}_u\right)+{\tilde \Phi_h}\mathbf{G}\cdot \mathbf{I}^{\alpha_u\beta_u}_u\right)u_R + {\rm H.c.},
\end{aligned}
\end{equation}
where $q_{R}$ $(q=u,d)$ is a  right-handed quark singlet, $Q_L$ is a left-handed $SU(2)$ quark doublet, ${F}$ and ${G}$ are general complex $4\times 4$ Yukawa  matrices in flavor space, $\mathbf{I}$ is the   $4\times 4$  identity matrix and $\mathbf{I}^{\alpha_q\beta_q}_q$ are diagonal  matrices defined as $\mathbf{I}^{\alpha_q\beta_q}_q={\rm diag}(0,0,\alpha_q,\beta_q)$. The Higgs doublets can be written as
\begin{equation}
\Phi_i=\begin{pmatrix}
      \phi_i^+    \\
      \frac{\upsilon_i+\phi^0_i}{\sqrt{2}}
\end{pmatrix},
\end{equation}
with $\tilde{\Phi}_i=i\sigma^2 \Phi_i$ ($i=\ell,\,h$).

The 4GTHDM is a variation of type-II THDM, therefore  the Yukawa Lagrangian  \eqref{Yukawa} has a  ${Z}_2$ symmetry, with the fields transforming as shown in Table \ref{Z2charges}.

\begin{table}[hbt!]
\caption{${Z}_2$ charges for the Higgs doublets and quarks in the 4GTHDM \cite{4G2HDM}.\label{Z2charges}}
\begin{tabular}{ccccccc}
  \hline
Field&$\Phi_\ell$&$\Phi_h$&$Q_L$&$d_R$, $s_R$, $u_R$, $c_R$&$t_R$, $b_R$&$t'_R$, $b'_R$\\
  \hline
${Z}_2$ charge&$-$&$+$&$+$&$-$&$(-1)^{1+\alpha_{t,b}} $&$(-1)^{1+\beta_{b',t'}} $\\
  \hline
\end{tabular}
\end{table}

The fermions of the fourth family can get their masses via the following three scenarios \cite{4G2HDM}:
\begin{itemize}
\item[(i)]($\alpha_{b}$, $\beta_{b'}$, $\alpha_{t}$, $\beta_{t'}$)=(0,1,0,1): $\Phi _{h}$ gives masses to the fermions of the fourth family only, whereas $\Phi_{\ell}$ gives masses to the remaining fermions.
\item[(ii)]($\alpha_{b}$, $\beta_{b'}$, $\alpha_{t}$, $\beta_{t'}$)=(1,1,1,1): $\Phi_{h}$ generates the masses of both the third and fourth families, whereas $\Phi_{\ell}$ generates the masses for all  other families.
\item[(iii)]($\alpha_{b}$, $\beta_{b'}$, $\alpha_{t}$, $\beta_{t'}$)=(0,1,1,1): $\Phi_{h}$ only couples to the fermions with masses at the electroweak scale.
\end{itemize}
In this work we only consider the case (i), which is still compatible with the LHC data on the 125 GeV Higgs boson \cite{Review4G2HDM, 1254G2HDM}.

The physical fields  $H^\pm$,  $h^0$, $H^0$, $A^0$ (it is customary to assume that $h^0$  is lighter than  $H^0$) are obtained after the diagonalization of the neutral and charged Higgs mass matrices:
\begin{equation}
\begin{aligned}
    & \phi_\ell^+=c_\beta G^+-s_\beta H^+,\\
    &\phi_h^-=s_\beta G^++c_\beta H^+,  \\
    &  \phi^0_{\ell}=c_\alpha H^0-s_\alpha h^0+i\left(c_\beta G^0-s_\beta A^0\right),\\
    &\phi^0_{h}=s_\alpha H^0+c_\alpha h^0+i\left(s_\beta G^0+c_\beta A^0\right) ,
\end{aligned}
\end{equation}
where $G^+$ and $G^0$ are the charged and neutral Goldstone bosons, $\alpha$ is the mixing angle in the $CP$-even neutral Higgs sector. From now on we use the  shorthand notation $c_a\equiv\cos a$, $s_a\equiv\sin a$, $t_a=\tan a$, for any angle $a$. 

In this model,  FCNCs arise at the tree level in the scalar sector. After introducing the mass eigenstates, the Yukawa interactions   can be written as   \cite{4G2HDM}
\begin{equation}
\mathcal{L}=\frac{g}{2m_{W}} f^\phi\bar{q}_{i}\left({S}_{ij}^{\phi}
 + {P}_{ij}^{\phi} \gamma_{5} \right)
q_{j}\phi+\rm{H.c.},
\label{Lagragian1}
\end{equation}
where $\phi=h^0$, $H^0$, $A^0$ and $H^\pm$.  For the neutral scalar bosons, the subscripts $i$ and $j$ run over  up or down quarks, whereas for the charged scalar boson $H^+$ $i$ ($j$) runs over up (down) quarks.  The  coupling constants $f^\phi$, ${S}_{ij}^{\phi}$, and ${P}_{ij}^{\phi}$ depend on the model parameters and are shown in  Table  \ref{Couplings}. In general ${S}_{ij}^{\phi}$ and ${P}_{ij}^{\phi}$ are given in terms of the complex entries of the $4\times4$ CKM matrix elements $U_{ij}$ and the  mixing matrix elements $\Sigma^{u,d} _{ij}$.  In the scenario i) described above,  the matrices ${\Sigma}^{u,d}$ are given as \cite{Review4G2HDM}
\begin{equation}
\label{mezclama}
\begin{aligned}
    &\Sigma^d_{ij}=\Sigma^d_{ij}\left(0,1,D_R\right)=D^\ast_{R,4i}D_{R,4j} ,  \\
    &\Sigma^u_{ij}=\Sigma^u_{ij}\left(0,1,U_R\right)=U^\ast_{R,4i}U_{R,4j}  ,
\end{aligned}
\end{equation}
where $D_R$ and $U_R$ are the unitary rotation matrices that diagonalize the quark mass matrix. Note that ${\Sigma}^d$ and ${\Sigma}^u$ depend on the elements of the fourth row of $D_R$ and $U_R$, respectively. Since  $D_{R,4i}$ and $U_{R,4i}$ parametrize the mixings between the quarks of the fourth-generation and those of the first three generations,  $\Sigma^d_{ij}$ and   $\Sigma^u_{ij}$ ($i,j=1,2,3$) are expected to be very small. This fact becomes evident in the parametrization  introduced in \cite{CKMb9}  in terms of one complex parameter $\epsilon_b=|\sin\theta_{bb'}|e^{i\delta_b}$
\begin{equation}\label{Sigma_d}
\mathbf{\Sigma}^d\simeq \left(\begin{array}{c|c}
\begin{matrix}\mathbf{0}\end{matrix}&\begin{matrix}\mathbf{0}\end{matrix}\\
\hline
\begin{matrix}\mathbf{0}\end{matrix}&
\begin{matrix}
|\epsilon_b|^2& \epsilon_b^*\left(1-\frac{ |\epsilon_b|^2}{2}\right) \\
\epsilon_b\left(1-\frac{ |\epsilon_b|^2}{2}\right)&\left(1-\frac{ |\epsilon_b|^2}{2}\right)
\end{matrix}
\end{array}
\right),
\end{equation}
where $\mathbf{0}$ is the $2\times 2$ zero matrix. A similar expression for ${\Sigma}^u$ is given in terms of the complex parameter $\epsilon_t=|\sin\theta_{tt'}|e^{i\delta_t}$. Furthermore, we will assume below a similar parametrization for the mixing matrix of the lepton sector ${\Sigma}^\ell$, which will be given in terms of the parameter $\epsilon_\ell$.

\begin{table}[hbt!]
\caption{$ f^{\phi} $ constants along with the scalar ${S}_{ij}^{\phi}$ and pseudoscalar ${P}_{ij}^{\phi}$  couplings of the physical scalar bosons of the 4GTHDM. The subscripts $i$ and $j$ run over up or down quarks for neutral scalar bosons, whereas $i$ ($j$) runs over up (down) quarks for the charged scalar boson. Here $I_{q} $ is the weak isospin ($I_d=-\frac{1}{2}$, $I_u=\frac{1}{2}$), whereas $\Sigma_{ij}^{u,d}$ are elements of the new complex mixing matrix ${\Sigma}^{u,d}$, and $U_{ij}$ are elements of the $4\times4$ CKM matrix. In addition, $ f^\pm_{ij}=\frac{1}{2}\left(m_{q_{i}} \Sigma_{ij}^{q}\pm m_{q_{j}} \Sigma_{ji}^{q*}\right)$, with $q=u$ ($d$) for up (down) quarks, and $ h^\pm_{ij}=\frac{1}{2}(t_{\beta}+ \frac{1}{t_{\beta}} )(m_{u_{k}} \Sigma_{ki}^{ u\ast}U_{kj}\pm m_{d_{k}} \Sigma_{kj}^{d}U_{ik})$. {We use the shorthand notation $s_a\equiv \sin a$,  $c_a\equiv \cos a$, and  $t_a\equiv \tan a$ ($a=\alpha,\,\beta$).}}
\label{Couplings}
\begin{center}
\begin{tabular}{cccccc}
\hline
\hline
$\phi$&  $ f^{\phi} $  &${S}_{ij}^{\phi}$ & ${P}_{ij}^{\phi} $ \\
\hline
\hline
$ h^{0} $& $\frac{c_{\alpha}} {s_{\beta}} +\frac{s_{\alpha}} {c_{\beta}} $  &
$\frac{m_{q_i}} {f^\phi}\frac{s_\alpha}{c_\beta}\delta_{ij}- f^+_{ij}$ &
$ -f^-_{ij}$ \\
\hline
$ H^{0} $& $\frac{c_{\alpha}} {c_{\beta}} -\frac{s_{\alpha}} {s_{\beta}} $
&$-\frac{m_{q_i}} {f^\phi}\frac{c_\alpha}{c_\beta}\delta_{ij}+ f^+_{ij}$ &
$ f^-_{ij}$ \\
\hline
$ A^{0} $ & $ 2iI_{q}(t_{\beta}+ \frac{1}{t_{\beta}} ) $&
$ f^- _{ij}$
&$-\frac{m_{q_i}} {f^\phi}t_\beta\delta_{ij}+ f^+_{ij}$  \\
\hline
$ H^{\pm} $&  $ \frac{2}{\sqrt{2}} $&$\frac{1}{2}t_{\beta}U_{ij}(m_{d_{j}} -m_{u_{i}} ) + h^-_{ij}$ &
$\frac{1}{2}t_{\beta}U_{ij}(m_{d_{j}} +m_{u_{i}} )- h^+_{ij}$
\\
\hline
\hline
\end{tabular}
\end{center}
\end{table}

In the alignment limit, which is given  by

\begin{equation}
c_{\beta-\alpha}\equiv \cos(\beta-\alpha)=0,
\end{equation}
the $h^0$ couplings to the SM particles are identical to those of the SM Higgs boson. So, it is natural to use as free parameters $t_\beta$ and $c_{\beta-\alpha}$.

\begin{figure}[htb!]
\begin{center}
\includegraphics[width=8cm]{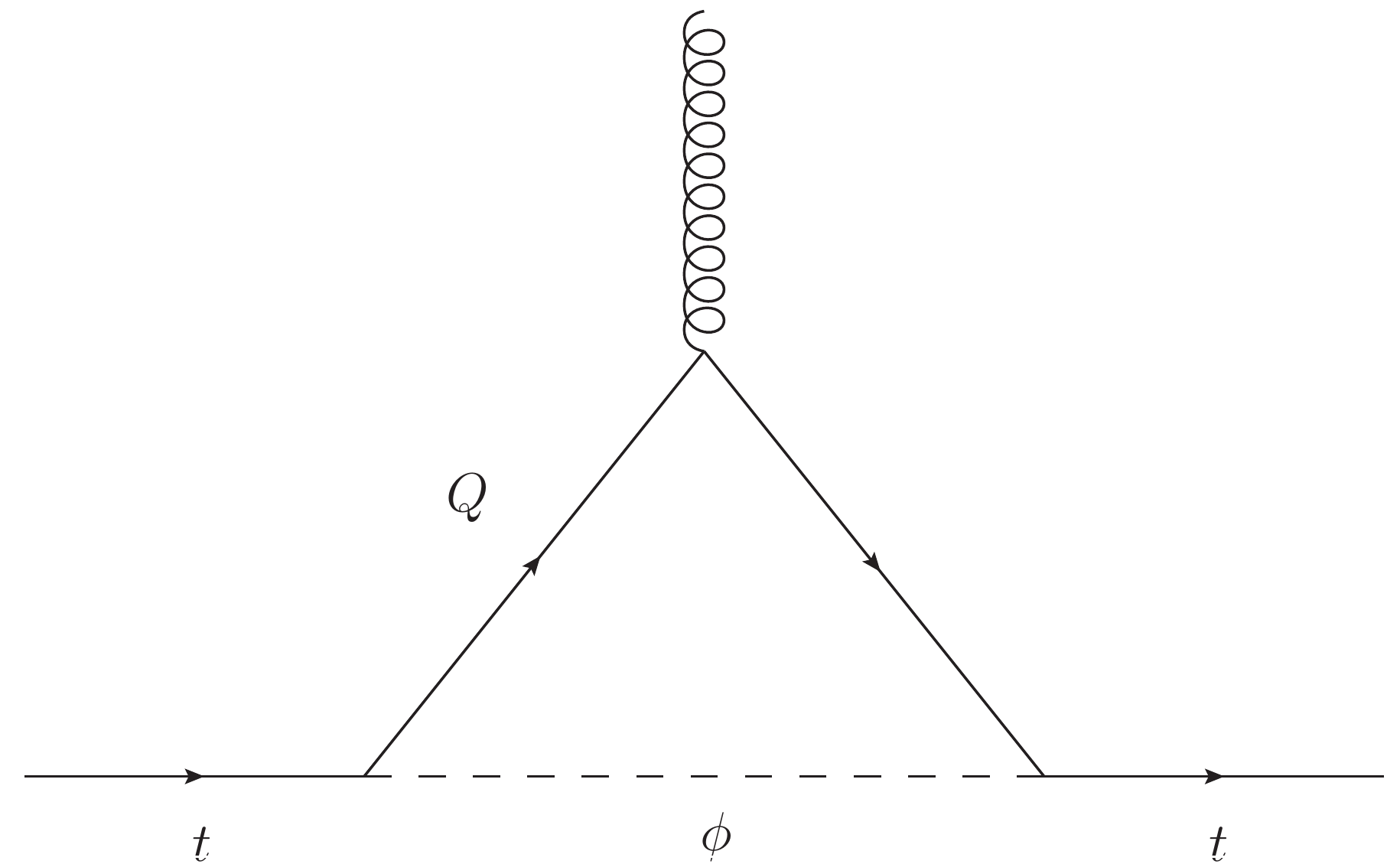}
\caption{One loop contribution to the CMDM and CEDM of the top quark in the 4GTHDM, where $Q=u_i$ for the neutral scalar bosons $\phi=H^0$ and $A^0$, whereas $Q=d_i$ for the charged scalar boson $\phi=H^\pm$, with $i=1\ldots 4$. In our calculation however we will only consider the contributions of the quarks of the third and fourth families.}
\label{FeynmanDiagram}
\end{center}
\end{figure}

\section{chromodipole moments of the top quark in the 4GTHDM }
\label{CromoDipoleMoments}
The most relevant contributions to the CMDM and CEDM of the top quark arise from the heaviest quarks, thus we only consider the contributions from the quarks of the third and fourth families. From the general Lagrangian (\ref{Lagragian1}), one can deduce that the one-loop level  scalar boson contributions to the CMDM  and CEDM of the top quark arise through the generic Feynman diagram of Fig. \ref{FeynmanDiagram}, where $Q=t, t^\prime$ for the neutral scalar bosons, whereas $Q=b, b^\prime$ for the charged scalar boson. After writing out the corresponding invariant amplitude for the  $\bar{t}t g$ vertex,  we have used both the Feynman parameter technique and the Passarino-Veltman reduction scheme to solve the loop integrals, which turn out to be free of ultraviolet divergences. We thus  write the contribution of the Feynman diagram of Fig. \ref{FeynmanDiagram} to the top quark CMDM and CEDM as follows:
\begin{eqnarray}
a_{t}^{\phi}(m_Q) &=&\left(\frac{g}{2 r_{W}} \right)^{2}\frac{|f_{\phi}|^{2} }{8\pi^{2}} \left( | {\tilde S}_{tQ}^\phi | ^2 F(r_Q,r_\phi)+| {\tilde P}_{tQ}^\phi | ^2 F(-r_Q,r_\phi)\right)
,\label{CMDM}
\\
d_t^\phi(m_Q) &=&\frac{g_s}{m_t}\left(\frac{g}{2 r_{W}} \right)^{2}\frac{|f_{\phi}|^{2}} {8\pi^{2}} \text{Im}\big({\tilde S}_{tQ}^\phi {\tilde P}_{tQ}^{\phi \ast} \big)  G(r_Q,r_\phi)\label{CEDM},
\end{eqnarray}
where for convenience we introduce the dimensionless parameters $r_{a}=m_{a}/m_t$, $\tilde S_{ij}^\phi=S_{ij}^\phi/m_t$, and $\tilde P_{ij}^\phi=P_{ij}^\phi/m_t$. Our result is consistent as the CEDM  requires a complex phase to be nonvanishing, whereas a nonzero CMDM does not require such a phase. The $F(x,y)$ and $G(x,y)$ functions are given in terms of Feynman parameter integrals as follows
\begin{equation}
\label{CMDMintFP}
F(x,y)=\int _{0}^{1} dz \frac{(1-z)^{2}(z+ x)}
{(1-z) (x^2-z)+z y^2},
\end{equation}
and
\begin{equation}
\label{CEDMintFP}
G(x,y)=x \int _{0}^{1}dz \frac{(1-z)^{2}} {(1-z) (x^2-z)+zy^2},
\end{equation}
for which explicit solutions are presented in Appendix \ref{asymptotic}, whereas the respective expressions  in terms of Passarino-Veltman scalar functions, obtained with the help of the FeynCalc package  \cite{Shtabovenko:2016sxi}, are given as follows
\begin{align}
\label{CMDMPV}
F(x,y)&=\frac{1}{2\delta^-_{xy}} \Big(2 y^2\left(y^2-x(x-1)\right)
   \Delta_y(x,y)-2 x
   \left(x\left(y^2-x(x-1)+1\right)-1\right) \Delta_x(x,y)\nonumber\\&+   \left(y^2-x^2\right) \left(2 y^2-2
   \left(x-1\right) x+1\right)+1\Big),
\end{align}
where $\delta^\pm_{xy}=y^2-\left(x\pm1\right)^2$ and
\begin{align}
\label{CEDMPV}
G(x,y)& =\frac{x}{\delta^-_{xy}\delta^+_{xy}} \Big(2 x y^2 \left(y^2-x^2+1\right) \Delta_y(x,y)+2 x
   \left(\left(x^2-1\right)^2-\left(y^2+1\right) y^2\right)
   \Delta_x(x,y)\nonumber\\&+2x \left(y^2-x^2+1\right)^2\Big),
\end{align}
where $\Delta_z(x,y)=B_0(0,m_t^2 z^2,m_t^2 z^2)-B_0(m_t^2,m_t^2 x^2,m_t^2 y^2)$, with $B_0(a,b,c)$ being two-point Passarino-Veltman scalar functions written as usually.  These alternative expressions are useful to cross-check the numerical results. Furthermore, in Appendix \ref{asymptotic} we present  closed expressions for the $F(x,y)$ and $G(x,y)$ functions, and analyze their asymptotic behavior at $x\gg y$, namely, for an ultra-heavy fourth-generation quark, which are useful to analyze the decoupling properties of the CMDM and CEDM.

To obtain the total contribution of the 4GTHDM to $a_t$, we must sum over all the scalar bosons, along with  the third- and fourth-generation quarks. However, as discussed below, $d_t$ only receives contribution from the charged scalar boson. It is  worth noting that since $c_{\beta-\alpha} \ll 1$, the contribution to the top quark CMDM from the loop with the lightest neutral Higgs boson $h^0$ and the top quark does not deviate considerably from that of the SM Higgs boson $h^0_{\rm SM}$, which follows straightforwardly from Eq. (\ref{CMDM}) after  substituting $\phi\to h^0_{\rm SM}$, $r_Q=1$, $f_{\phi}=S^\phi_{tt}=1$, and $P^\phi_{tt}=0$:
 \begin{equation}
\label{hinduss}
a_t^{h^0_{\rm SM}} =\frac{G_F m_t^2}{4\sqrt{2}\pi^2}\int^1_0 dz \frac{(1+z)(1-z)^2}{(1-z)^2+z r_{h^0_{\rm SM}} ^2},
\end{equation}
which agrees with results reported  previously in the literature \cite{ManyModels,CMDMenSM} and is also in accordance with the corresponding contribution to the top quark anomalous MDM.
By using  $m_t=173$ GeV and $m_{h^0_{\rm SM}} =125$ GeV, we can obtain the following numerical value
\begin{equation}
\label{atSM}
a^{h^0_{\rm SM}} _t=3.78\times 10^{-3}
\end{equation}
As $a_q^{h^0_{\rm SM}} $ is  proportional to $m_q^2/m_W^2$, the CMDMs of light quarks are  considerably suppressed,  thus the top quark offers the best opportunity to study this property.

\section{Numerical analysis and discussion}
We now  analyze   the parameter space of the 4GTHDM and the most up-to-date constraints from experimental data.

\label{Numerical}
\subsection{Constraints and parameter space of the 4GTHDM }\label{sec:3}
According to the results given in Eqs. (\ref{CMDM}) and (\ref{CEDM}) along with Table \ref{Couplings}, we need the following parameters for our calculation: $t_\beta$, $c_{\beta-\alpha}$, the masses of the heavy scalar bosons and  the fourth-generation quarks, the $4\times 4$ CKM matrix elements $U_{ij}$ ($i=t,t'$ and $j=b,b'$), and the mixing matrix  elements $\Sigma_{ij}^u$ and  $\Sigma_{ij}^d$  ($i,j=3,4$). We turn to discuss the constraints on these parameters from current experimental data.

\subsubsection{Masses of the fourth-generation quarks}

The ATLAS and CMS collaborations have searched for signals of heavy quarks $Q$ at the LHC via  pair production $pp\to \bar{Q}Q$, though their results are  model dependent and focus mainly on vectorlike quarks, which  do not contribute to Higgs boson production via gluon fusion, thereby being compatible with LHC data. Such analyses  assume that vectorlike quarks with  SM-like electric charges  decay dominantly into one of the following channels $B\to Wt$, $Zb$, or $Hb$, for a charge $-1/3$ quark,  and $T\to Wb$, $Zt$, or $Ht$, for a charge $2/3$ quark \cite{Patrignani:2016xqp}.
Such searches have also been used to constraint the masses of  new chiral quarks. In particular the ATLAS collaboration found that new chiral $b'$ quarks with masses below 730 GeV are excluded at 95\% C.L. if  $b'\to tW$ is assumed to be the main decay channel with a 100\% branching ratio, but such a limit is considerably  relaxed, up to around  $400$ GeV, when one assumes that BR$(b'\to cW)\sim 1$ \cite{Aad:2015gdg}. Another recent report by the ATLAS collaboration \cite{Aad:2015tba} focuses on the search for pair production of  vectorlike and fourth-generation chiral quarks $Q$ decaying exclusively as $Q\to Wq$, $Zq$, and $hq$ $(q=u,d,s)$.  It was found that new chiral quarks with masses below 690 GeV are excluded at 95\% C.L. provided that BR$(Q\to qW)\sim 1$. Following Ref. \cite{Bar-Shalom:2016ehq}, for our analysis below we will assume that  the decays $Q\to Wq$, $Zq$, and $hq$ are suppressed due to suppressed values of the mixing matrices $ \Sigma^q$ and the $4\times 4$ CKM matrix $U$. In this way,  the  lower bounds on the fourth-generation quark masses could be evaded and one can consider much lighter $b'$ and $t'$ quarks.   As for the $m_{b',t'}$ upper values, unitarity constrains the mass of a chiral quark doublet around 550 GeV  \cite{Chanowitz:1978uj,Chanowitz:1978mv}. Furthermore, since the 4GTHDM is inspired in the idea that it is  the low-energy effective limit of a strongly interacting theory valid up to the TeV scale \cite{4G2HDM}, we will refrain from considering the scenario with  ultra-heavy quarks as unknown nonperturbative effects could turn relevant, thereby rendering our calculation unreliable. We will thus consider the interval 350--600 GeV for the fourth-generation quark masses, which was used in Ref. \cite{1254G2HDM}, where  the 4GTHDM parameter space was analyzed in view of the LHC data on the 125 Higgs boson.  In addition, we will see below that the mass splitting of the fourth-generation quarks is restricted  by the constraints on the oblique parameters $S$ and $T$ \cite{Review4G2HDM}.

{
\subsubsection{Mixing angles $t_\beta$ and $c_{\beta-\alpha}$}
 Since the 4GTHDM lightest scalar boson  $h^0$ must be identified with the scalar particle discovered at the LHC, whose properties are compatible with those of the SM Higgs boson, the $h^0$ couplings are not allowed to deviate considerably from those of the SM Higgs boson. The fit on  the Higgs boson coupling modifiers   $\kappa_i^2=\Gamma(h^0\to i)/\Gamma(h^0_{\rm SM}\to i)$ obtained from the combined data of the ATLAS and CMS collaborations at $\sqrt{s}=7$ and 8 TeV \cite{Khachatryan:2016vau}  can place constraints on the 4GTHDM parameters $t_\beta$, $c_{\beta-\alpha}$, $\epsilon_q$, $\epsilon_\ell$  and the masses of the fermions of the fourth generation.
 We have found that the coupling modifiers $\kappa_i$  are highly sensitive to  $t_\beta$, $c_{\beta-\alpha}$, and $|\epsilon_t|$, and  thus fix the values of the remaining parameters as follows $m_{b'}=350$ GeV, $m_{t'}=450$ GeV, $|\epsilon_b|\simeq O(m_b/m_{b'})\simeq 0.01$, $|\epsilon_\ell|= 0.1$, $m_{\nu'}=300$ GeV, and $m_{\ell'}=400$ GeV. We then show in Fig.  \ref{BoundsHiggsdata}   the allowed areas at 95 \% C.L. in the $t_\beta$ vs $c_{\beta-\alpha}$ plane (top plots)   and $t_\beta$ vs $|\epsilon_t|$ plane (bottom plots) consistent with the constraints on $\kappa_W$, $\kappa_Z$, $\kappa_t$, $\kappa_b$, $\kappa_\tau$, $\kappa_\gamma$, and $\kappa_g$ \cite{Khachatryan:2016vau}. For the Higgs boson coupling modifiers we have implemented our own code with $\kappa_f=s_\alpha/c_\beta$ ($f=\ell,\, q_u,\, q_d$) and $\kappa_V=s_{\beta-\alpha}$ ($V=W,\, Z$) for the tree-level couplings, whereas for  the one-loop induced coupling modifiers $\kappa_\gamma$ and $\kappa_g$  we use the formulas reported in  \cite{Gunion:1989we,Djouadi:2005gj} for the leading contributions to the decays of a $CP$-even Higgs boson into photon and gluon pairs, including the  contributions of a fourth generation of fermions (the contribution of the charged Higgs scalar boson to $\Gamma(h\to \gamma\gamma)$ is negligible). We thus have:
\setcounter{equation}{16}
\begin{equation}
\kappa_\gamma^2\simeq \dfrac{\left|\kappa_W F_1\left(\tau_W\right) +\sum\limits_{f=t,t',b',\ell'}\kappa_f N_f Q_f^2 F_{1/2}\left(\tau_f\right)\right|^2}{\left|F_1\left(\tau_W\right) +\frac{4}{3}F_{1/2}\left(\tau_t\right)\right|^2},
\end{equation}
with   $\tau_a\equiv 4 m_{h^0_{\rm SM}}^2/m_a^2$, $N_f$ being the fermion color number,  and
\begin{equation}
F_{s}(\tau_a)=\left\{
\begin{array}{lcl}
-2\tau_a(1+(1-\tau_a)f(\tau_a)) &&\quad s=1/2,\\ \\
2+3\tau_a+3\tau_a(2-\tau_a)f(\tau_a)&&\quad s=1,
 \end{array}\right.
\end{equation}
where the $f(x)$ function is given by
\begin{equation}
f(x)=\left\{
\begin{array}{cr}
\left[\arcsin\left(\frac{1}{\sqrt{x}}\right)\right]^2&x\ge1,\\
-\frac{1}{4}\left[\log\left(\frac{1+\sqrt{1-x}}{1-\sqrt{1-x}}\right)-i\pi\right]^2&x<1.
\end{array}
\right.
\end{equation}
In addition, for $\kappa_g$ we have
\begin{equation}
\kappa_g^2\simeq \dfrac{\left|\sum\limits_{q=t,t',b'}\kappa_q  F_{1/2}\left(\tau_q\right)\right|^2}
{\left|F_{1/2}\left(\tau_t\right)\right|^2},
\end{equation}
See also \cite{Das:2017mnu} for a similar treatment of the Higgs boson coupling modifiers within  a THDM with a fourth generation of fermions.

\begin{figure}[htb!]
  \centering
  \includegraphics[width=14cm]{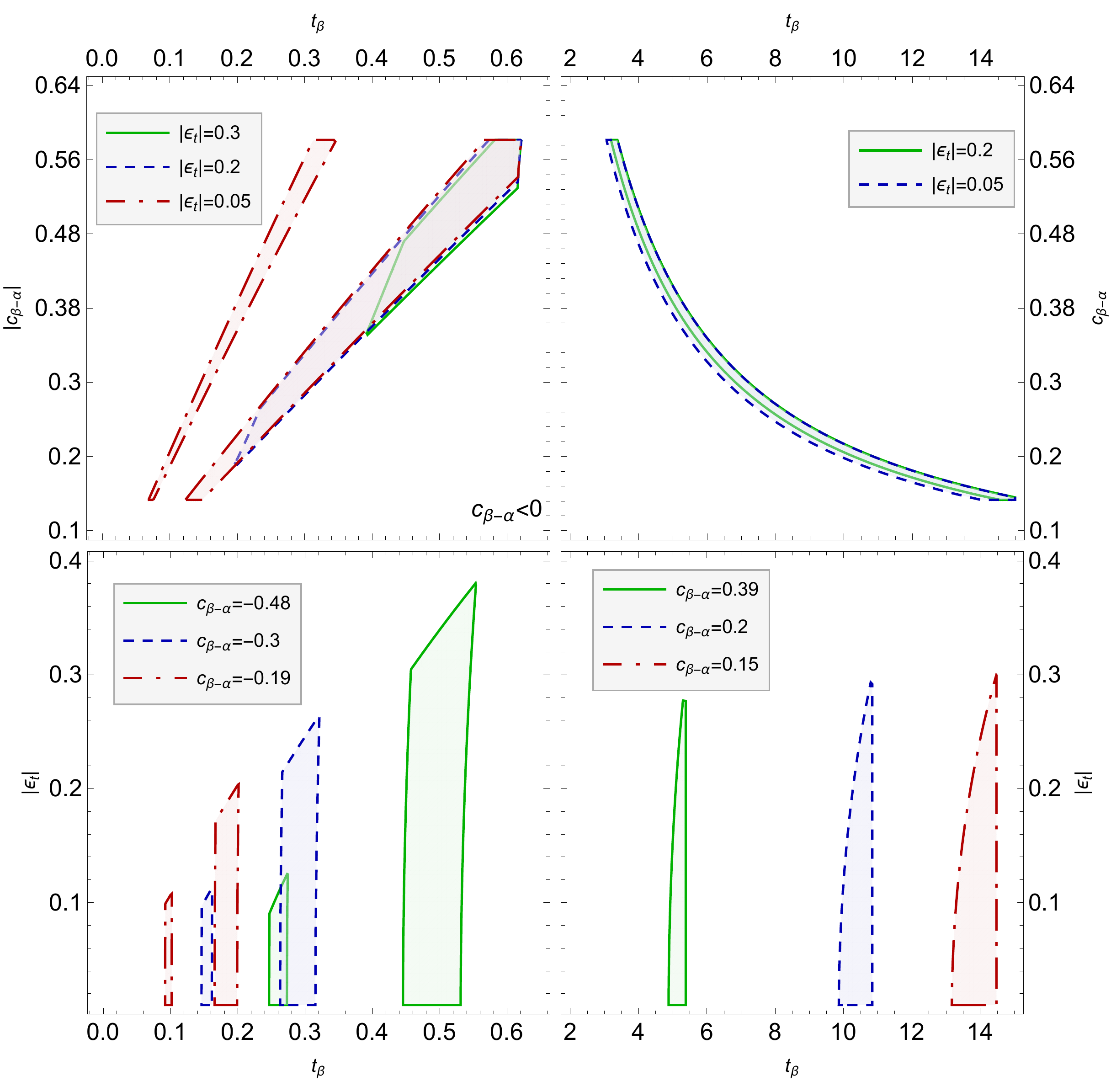}
  \caption{Allowed areas at 95\% C.L.  consistent with the fit on the coupling modifiers $\kappa_i$ obtained from the combined $\sqrt{s}=7$ and 8 TeV data of the ATLAS and CMS collaborations  \cite{Khachatryan:2016vau}.  The top plots show the allowed area in the  $t_\beta$ vs $c_{\beta-\alpha}$ plane for several values of $|\epsilon_t|$,  and  the bottom plots show the allowed area in the $t_\beta$ vs $|\epsilon_t|$ plane for  distinct values of $c_{\beta-\alpha}$. We have fixed the values of the remaining parameters of the model as follows $m_{b'}=350$ GeV, $m_{t'}=450$ GeV, $|\epsilon_b|\simeq O(m_b/m_{b'})\simeq 0.01$, $|\epsilon_\ell|= 0.1$, $m_{\nu'}=300$ GeV, and $m_{\ell'}=400$ GeV.}\label{BoundsHiggsdata}
\end{figure}

In the top plots of Fig. \ref{BoundsHiggsdata} we observe that there are two regions consistent with the constraints on the Higgs coupling modifiers: in the first scenario (top-left plot) the allowed area lies within $0.05\lesssim t_\beta\lesssim 0.6$ and $-0.59\lesssim c_{\beta-\alpha}\lesssim -0.16$, depending on the value of $|\epsilon_t|$, whereas in the second scenario (top-right plot)  $c_{\beta-\alpha}$ is constrained to lie in the interval from $0.16$ to $0.58$, while $3\lesssim t_\beta\lesssim  15$.  These results are in agreement with those found in Ref. \cite{Bar-Shalom:2016ehq}.
We observe that for $t_\beta<0.6$, the allowed area  reduces considerably for smaller $|c_{\beta-\alpha}|$. This is evident in the allowed area in the $t_\beta$ vs $\epsilon_t$ plane (bottom left plot): when $c_{\beta-\alpha}=-0.19$,  the allowed area lies within  two short narrow  bands centered around $t_\beta\simeq 0.1$, where $|\epsilon_t|\le 0.1$,  and $t_\beta\simeq 0.19$, where $|\epsilon_t|\le 0.2$. We observe that the height and width of the allowed bands increases as $|c_{\beta-\alpha}|$ increases.
We also note  that for $t_\beta>3$, smaller values of $t_\beta$ require larger values of $c_{\beta-\alpha}$ and vice versa, and the allowed area shrinks significantly if  $|\epsilon_t|$  increases  by one order of magnitude. For instance, for $c_{\beta-\alpha}=0.2$, the allowed band in the $t_\beta$ vs $|\epsilon_t|$ plane shrinks significantly as $|\epsilon_t|$ increases (bottom-right plot): for small $|\epsilon_t|$, $t_\beta$ is constrained to lie within a narrow band between 10 and 11, but the width of such a band shrinks considerably as $|\epsilon_t|$ increases up to 0.3.
Below we consider the following two set of values consistent with the constraints on the Higgs coupling modifiers: $(t_\beta,c_{\beta-\alpha})=(10,0.19)$ and $(t_\beta,c_{\beta-\alpha})=(5,0.40)$, dubbed scenarios I and II, respectively, from now on.

}

\subsubsection{Masses of the heavy scalar bosons}

{
The existence of new  scalar bosons has been explored by the ATLAS and CMS Collaborations: a  heavy scalar boson $H^0$ has been searched for in  the  $\gamma\gamma$ \cite{Khachatryan:2015qba}, $ZZ$ \cite{Aad:2015kna}, $h^0h^0$ \cite{Khachatryan:2015tha} and $\tau\tau$ \cite{Aaboud:2016cre, Aaboud:2017sjh} channels, whereas the pseudoscalar boson $A^0$ has been looked for in the $\gamma\gamma$ \cite{Khachatryan:2015qba}, $Zh^0$ \cite{Aad:2015wra, Khachatryan:2015lba, Khachatryan:2015tha} and $\tau\tau$ \cite{Aaboud:2016cre, Aaboud:2017sjh} channels.}  The corresponding bounds are model dependent, focusing mainly on the MSSM and THDMs. Along these lines, the ATLAS collaboration used  the LHC data at $\sqrt{s}=8$ TeV  on the  $H\to ZZ$ channel to search for a heavy neutral scalar boson and their results were interpreted in type-I and type-II THDMs \cite{Aad:2015kna}. As for the  type-II THDM, a $CP$-even Higgs boson with mass $m_H=200$ GeV was considered and the exclusion region in the $t_\beta$ vs $c_{\beta-\alpha}$ plane was found: for $t_\beta<1$, only a very narrow  area centered around $c_{\beta-\alpha}\sim 0$ is still allowed, but  for $t_\beta>2$ the allowed region expands considerably, so that values up to $c_{\beta-\alpha}\sim 0.6$ are still allowed. Similar constraints were found for a $CP$-odd scalar boson, which was searched for using the $A\to h Z$ channel by the ATLAS  \cite{Aad:2015wra} and CMS collaborations \cite{Khachatryan:2015lba}, complemented with the search via  the $A\to \bar{\tau}\tau$ channel \cite{Aad:2014vgg}: it was  found that for $m_A=300$ GeV, the region with $t_ \beta\lesssim 2$ is forbidden for any $c_{\beta-\alpha}$, but there is a wide area with $t_\beta>2$ and $-0.2\lesssim c_{\beta-\alpha}\lesssim 0.4$ still allowed. As for the charged scalar boson the direct search at LEP imposed the constraint $m_{H^\pm} > 80$ GeV \cite{Abbiendi:2013hk}, but the search at the LHC is challenging as the QCD background is very high: a charged scalar boson was searched for  through the decays $t\rightarrow H^\pm b$ and $H^\pm\rightarrow \tau^+\nu_\tau$ \cite{Khachatryan:2015qxa}, though the results were interpreted in the context of the MSSM. There are also indirect constraints on the mass of the charged scalar boson in the context of THDMs, which can be obtained through the bounds on the experimental measurements on the $Z\to \bar{b}b$ decay and low energy FCNC processes. It turns out that the measurement of the $\bar B \to X_s\gamma$ branching ratio imposes the stringent lower bound  $m_{H^\pm} > 570$ GeV, independently of $t_\beta$, in the usual type-II THDM\cite{Misiak:2017bgg}.

Although the above bounds are not directly applicable to the 4GTHDM, we expect  no considerable deviation  in the limit of  $|U_{t'b}|\ll 1$, $|U_{b't}|\ll 1$, and small mixing between the fourth-generation fermions and the SM ones, in which the usual type-II THDM is recovered, so we will consider  scalar boson masses for the neutral scalar bosons above 400 GeV, whereas for the charged scalar boson mass we use values  above 600 GeV, unless stated otherwise.
Constraints from direct searches can be complemented with those  obtained from  vacuum stability and unitarity of the scalar potential along with perturbativity of the Higgs couplings. However, we do not take into account this class of constraints as  the  4GTHDM is an effective theory with unknown scalar potential (the underlying fundamental theory is  unknown). Even if an effective scalar potential is set up, its parameters would receive large radiative corrections from the UV completion of the theory \cite{Review4G2HDM}.

\begin{figure}[htb!]
  \centering
  \includegraphics[width=14cm]{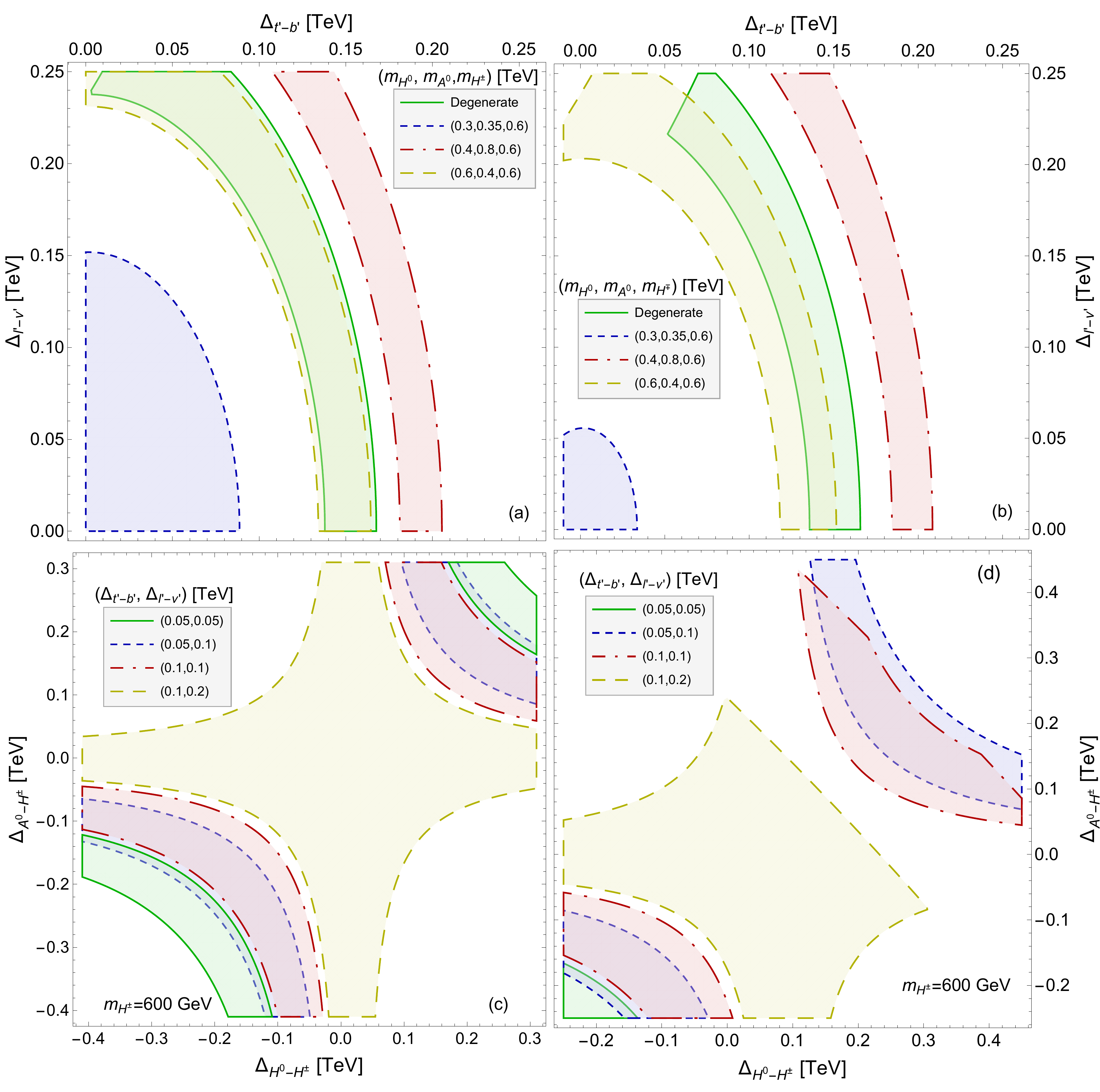}
  \caption{Allowed areas at 95\% C.L.  consistent with the constraints on the oblique parameters \cite{Patrignani:2016xqp} in the  $\Delta_{t'-b'}$ vs $\Delta_{\ell'-\nu'}$   plane for four sets of $(m_{H^0},m_{A^0},m_{H^\pm})$ values (top plots), and the $\Delta_{H^0-H^\pm}$ vs $\Delta_{A^0-H^\pm}$ plane for  a few sets of $(\Delta_{t'-b'},\Delta_{\ell'-\nu'})$ values and  $m_{H^\pm}=600$ GeV (bottom plots). In the left (right) plots we use $(t_\beta,c_{\beta-\alpha})=(10,0.19)$ [$(t_\beta,c_{\beta-\alpha})=(5,0.4)$].  We also use $m_{\nu'}=300$ GeV and $m_{b'}=350$ GeV. }\label{BoundsOblique}
\end{figure}

Other constraints arise from the oblique parameters $S$ and $T$, which  bound the mass splitting of the scalar bosons. Since  $S$ and $T$  also depend on the mass splitting of the quarks (leptons) of the fourth generation, a more careful analysis  is in order here. We first define the splitting between the masses of particles $A$ and $B$ as follows: $\Delta_{A-B}=m_A-m_B$. The analytical expressions of the oblique parameters necessary for our calculation can be found for instance in \cite{Haber:1999zh,He:2001tp,Eberhardt:2010bm,Dighe:2012dz,Bar-Shalom:2016ehq} (for completeness we present the corresponding expressions in Appendix \ref{oblique}). To obtain constraints on the mass splitting of the fourth-generation fermions from the bounds on the oblique parameters \cite{Patrignani:2016xqp} we find it convenient to fix  $m_{\nu'}=300$ GeV and $m_{b'}=350$ GeV. We then show in Fig. \ref{BoundsOblique} the allowed values of the mass splitting at 95\% C.L. in some illustrative scenarios. In the top plots we show the allowed areas in the $\Delta_{t'-b'}$ vs $\Delta_{\ell'-\nu'}$   plane for four sets of $(m_{H^0},m_{A^0},m_{H^\pm})$ values, whereas the bottom plots  show the allowed areas in the $\Delta_{H^0-H^\pm}$ vs $\Delta_{A^0-H^\pm}$ plane for  a few sets of $(\Delta_{t'-b'},\Delta_{\ell'-\nu'})$ values and  $m_{H^\pm}=600$ GeV. In the left plots we use $(t_\beta,c_{\beta-\alpha})=(10,0.19)$ and in the right plots we set $(t_\beta,c_{\beta-\alpha})=(5,0.4)$. We  highlight the following features than can be drawn from the analysis of these plots: i)the heavy scalar bosons can have degenerate masses provided that there is a nonzero $\Delta_{t'-b'}$ lying in the interval from 50 to 150 GeV, for $0\le \Delta_{\ell'-\nu'}\le $ 200 GeV; ii)both  $\Delta_{t'-b'}$ and $\Delta_{\ell'-\nu'}$ can be small or large as long as there is either nonzero $\Delta_{H^0-H^\pm}$ or nonzero $\Delta_{A^0-H^\pm}$;
iii) if $m_{H^0}$ ($m_{A^0}$) is relatively light, both $m_{H^\pm}$ and $m_{A^0}$ ($m_{H^0}$) can become simultaneously heavy. An interesting scenario arises  when $\Delta_{t'-b'}\sim 100$ GeV, $\Delta_{\ell'-\nu'}\sim 200$ GeV, and $m_{H^\pm}=600$ GeV (areas with long-dashed borders in the bottom plots) as  a wide range of values of the masses of the heavy scalar bosons are allowed, including  degenerate ones. However,  the allowed area is considerably larger for $(t_\beta,c_{\beta-\alpha})=(10,0.19)$ as shown in the bottom-left plot.  Below we consider values for the heavy scalar boson masses fulfilling these constraints.

\subsubsection{$U$ and $\Sigma^q$ matrix elements}\label{sec:elements}
The diagonal and nondiagonal elements of the $4\times 4$ matrices  ${U}$ and  ${\Sigma}^q$  are involved in our analysis, consequently a more detailed discussion is required.  We first write the corresponding matrix elements in exponential form
\begin{equation}
\label{euler}
 U_{ij}=|U_{ij}|e^{i\rho_{ij}} \quad\text{and}\quad \Sigma^q_{ij}=|\Sigma^q_{ij}|e^{i\eta^q_{ij}} ,
\end{equation}
and discuss the implications of unitarity and hermicity on the moduli and phases.

A  $4\times 4$ unitary matrix can be parameterized by six mixing angles and three $CP$-violating complex phases \cite{Hou:1987hm}, but  we only need the $U_{ij}$ ($i=t,t'$ and $j=b,b'$) elements for our analysis. For the diagonal elements,  $\rho_{ii}=0$ due to unitarity and we can assume $|U_{ii}|\simeq 1$. Furthermore, we can  take $|U_{t'b}|\simeq |U_{tb'}|$, $\rho_{t'b}\simeq 0$ and $\rho_{tb'}\ne0$ without losing generality \cite{Hou:1987hm}. Thus, $|U_{tb'}|$ and $\rho_{tb'}$ will be the only free parameters involved in  the CMDM and CEDM. From the experimental data on $Z$, $K$,  and $B$ decays as well as  $B$-meson mixing,  the upper bound $|U_{tb'}|<0.12$  was extracted \cite{CKMb9}. We will then use $|U_{tb'}| \simeq 10^{-1}$ and $\rho_{tb'}\in [-\pi,\pi]$.

As ${\Sigma}^q$ is Hermitian [see Eq. \eqref{mezclama}],  its diagonal elements must be real  ($\eta^q_{ii}=0$), whereas its nondiagonal elements must obey $|\Sigma^u_{ij}|=|\Sigma^u_{ji}|$ and $\eta^u_{ij}=-\eta^u_{ji}$. This leaves $|\Sigma^u_{33}|$, $|\Sigma^u_{34}|$,$|\Sigma^u_{44}|$ and $\eta^u_{34}$ as   free parameters, along with an identical number of free  parameters associated with the  $ \Sigma^d$ matrix. As explained above, these   matrices  parametrize the mixing between the fourth-generation quarks and those of the first three generations. Instead of the parametrization of Eq. \eqref{euler}, we will use the parametrization of Eq.  \eqref{Sigma_d} in terms of the complex parameters $\epsilon_t$ and $\epsilon_b$, which means that  $\eta^u_{43}\equiv \delta_t$ and   $\eta^d_{43}\equiv \delta_b$.
However, there are no  experimental bounds on these parameters, though the authors of \cite{CKMb9} considered the values $|\epsilon_b|\simeq O\left(\frac{m_b}{m_{b'}} \right)\simeq O(0.01)$ and $|\epsilon_{t}|\simeq O\left(\frac{m_t}{m_{t'}} \right)\simeq O(0.1)$, which we use in our analysis below.

\subsubsection{Summary of benchmarks used for the evaluation of the top quark CMDM and CEDM}
In conclusion, in our analysis we will   consider  two illustrative scenarios for the values of the parameters $t_\beta$ and $c_{\beta-\alpha}$, consistent with the LHC Higgs data:
\begin{enumerate}
\item[]Scenario I: $(t_\beta, c_{\beta-\alpha})=(10,0.19)$.
\item[]Scenario II: $(t_\beta, c_{\beta-\alpha})=(5,0.4)$.
\end{enumerate}
For the remaining parameters we use the values shown in Table   \ref{parameters},  focusing on  values of the heavy scalar boson masses consistent with the constraints discussed above.

\begin{table}[htb!]
\centering
\caption{Values used for the parameters of the 4GTHDM in the analysis of the top CMDM and CEDM, unless indicated otherwise. Here $\phi$ stands for the heavy scalar bosons. For the mixing matrix ${\Sigma}^u$, we adopt the parametrization of Eq. \eqref{Sigma_d}, with an analogue parametrization for ${\Sigma}^d$, and  two scenarios for the values of $t_\beta$ and $c_{\beta-\alpha}$ consistent with the LHC Higgs data: $(t_\beta, c_{\beta-\alpha})=(10,0.19)$ (scenario I) and  $(t_\beta, c_{\beta-\alpha})=(5,0.4)$ (scenario II).\label{parameters}}
\begin{tabular}{ccc}
 \hline\hline
 Parameter&&Value\\
 \hline\hline
 $m_{b'}$, $m_{t'}$  &&$350-600$  GeV\\
$\Delta_{t'-b'}$ && 120 GeV\\
 $m_\phi$  &&400--1000 GeV \\
 $|U_{tb}|$, $|U_{t'b'}|$&&$0.99$\\
 $|U_{t'b}|$, $|U_{tb'}|$&&$0.1$\\
 $\rho_{t'b}$&&$0$\\
 $|\epsilon_t|$, $|\epsilon_b|$&&$0.1$, $0.01$\\
 $\rho_{tb'},\delta_t, \delta_b$&&$\pi/2$, $\pi/4$, $\pi/4$\\
 \hline
\hline
\end{tabular}
\end{table}

\subsection{Top quark CMDM and CEDM in the 4GTHDM}
 For the  evaluation of the CMDM and CEDM of the top quark we use  the Mathematica routines for numerical  integration of Eqs. (\ref{CMDMintFP}) and (\ref{CEDMintFP}). A cross-check was done by evaluating the respective expressions in terms of Passarino-Veltman scalar functions [Eqs. (\ref{CMDMPV}) and (\ref{CEDMPV})] via the LoopTools routines \cite{Hahn:1998yk,vanOldenborgh:1989wn}.

\subsubsection{Top quark CMDM}
In the 4GTHDM there are new  contributions to the top quark CMDM arising from all the scalar bosons, but in our analysis we only consider the new physics contributions, so we remove the pure SM Higgs boson contribution given in Eq. \eqref{atSM}. The total contribution of the 4GTHDM is thus given as $a_t^{\rm 4GTHDM}=a^{\rm SM}_t+\delta a^{\rm 4GTHDM}_t$, where the new physics contribution $\delta a^{\rm 4GTHDM}_t$ is given as follows

\begin{equation}
\delta a_t^{\rm 4GTHDM}=a^{\rm 3rd}_t+a^{\rm 4th}_t,
\end{equation}
where $a^{\rm 3rd}_t$ and $a^{\rm 4th}_t$ are the contributions of the loops with internal quarks of the third and fourth generations, respectively, which can be written as

\begin{equation}
\label{at3rd}
a_t^{\rm 3rd}=\delta a^{h^0}_t(m_t)+\sum_{\phi=H^0,A^0}a^\phi(m_t)+a^{H^\pm}(m_b),
\end{equation}
and
\begin{equation}
\label{at4th}
a_t^{\rm 4th}=\sum_{\phi=h^0,H^0,A^0}a^\phi(m_{t'})+a^{H^\pm}(m_{b'}).
\end{equation}
with  $\delta a^{h^0}_t(m_t)=a^{h^0}_t(m_t)-a_t^{h^0_{\rm SM}} $ being the new physics correction to $a_t^{h^0}$ arising from the loop with  $h^0$ and $t$ quark exchange. Notice that in this model the $H^-\bar{b}t$ coupling depends on $m_{b'}$ and $m_{t'}$, so the third-generation quark contribution also depends on the masses of the fourth-generation quarks.

\begin{figure}[htb!]
  \centering
\includegraphics[width=17cm]{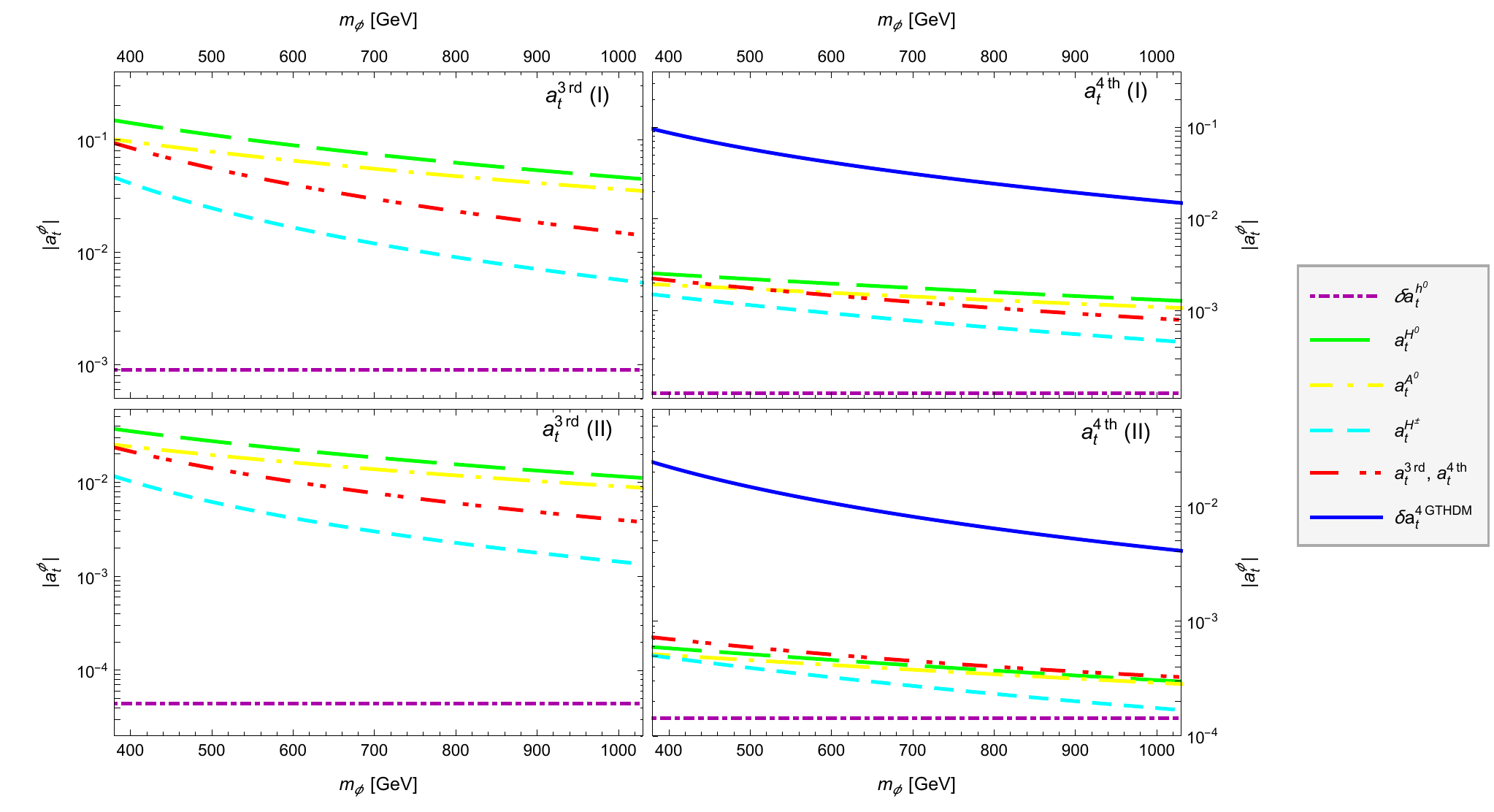}
  \caption{Partial contributions from the heavy scalar bosons of the 4GTHDM to $a_t$ as  functions of their masses, which are taken as degenerate,  for $(t_\beta,c_{\beta-\alpha})=(10,0.19)$ (top plots) and   $(t_\beta,c_{\beta-\alpha})=(5,0.4)$ (bottom plots). We show separately the partial contributions to $a_t^{\rm 3rd}$   (left plots) and $a_t^{\rm 4th}$ (right plots)  as well as the total contribution for each  generation (dash-dotted lines). The total new physics contribution $\delta a_t^{\rm 4GTHDM}$  is denoted by the solid lines in the right plots. We set $m_{b'}=350$ GeV  and for the remaining parameters we use the values shown in Table \ref{parameters}.}\label{atphidependence}
\end{figure}

We start  our analysis by assessing the impact of the presence of  the new heavy quarks on $a_t$ as they are the new ingredient of the 4GTHDM as compared to the usual  THDMs. We first assume that all the heavy scalar bosons have a degenerate mass $m_\phi$, which is allowed by the constraints on the Higgs coupling modifiers and the oblique parameters.  In Fig. \ref{atphidependence} we show  the behavior of the partial contributions of the light and heavy scalar bosons to $a_t$ as functions of $m_\phi$  for the  parameter values   of Table \ref{parameters}, with  $(t_\beta,c_{\beta-\alpha})=(10,0.19)$ (top plots) and $(t_\beta,c_{\beta-\alpha})=(5,0.4)$ (bottom plots). The left  plots show the partial  contributions to $a_t^{\rm 3rd}$ and the right plots those to $a_t^{\rm 4th}$ [Eqs. \eqref{at3rd} and \eqref{at4th}, respectively].  We observe that the main contributions to $\delta a_t^{\rm 4GTHDM}$ arise from the  loops including the heavy scalar bosons and the third-generation quarks ($a_t^{\rm 3rd}$), whereas all other contributions are subdominant, with the lightest Higgs boson giving the smallest contributions. In general,  all the  heavy scalar bosons give contributions of similar order of magnitude, though that of the charged scalar boson is slightly smaller.Therefore,  $\delta a_t^{\rm 4GTHDM}$ arises mainly from the loops with the heavy neutral scalar bosons accompanied by the top quark. However, due to their opposite signs there are cancellation between the distinct contributions, so $\delta a_t^{\rm 4GTHDM}$  is smaller than the partial contributions. We can conclude that  $\delta a_t^{\rm 4GTHDM}$ can reach values as large as $10^{-1}$  in the scenario with  $(t_\beta,c_{\beta-\alpha})=(5,0.4)$ and for relatively light  $m_\phi\sim $ 400--500 GeV.  We also would like to point out  that even in the limit $|U_{t'b}|\to 0$ and $|U_{tb'}|\to 0$,  $\delta a_t^{\rm 4GTHDM}$ remains unchanged as the charged Higgs boson contribution is subdominant. Also, there is no considerable effect arising from the fourth-generation fermions.

\begin{figure}[htb!]
  \centering
  \includegraphics[width=17cm]{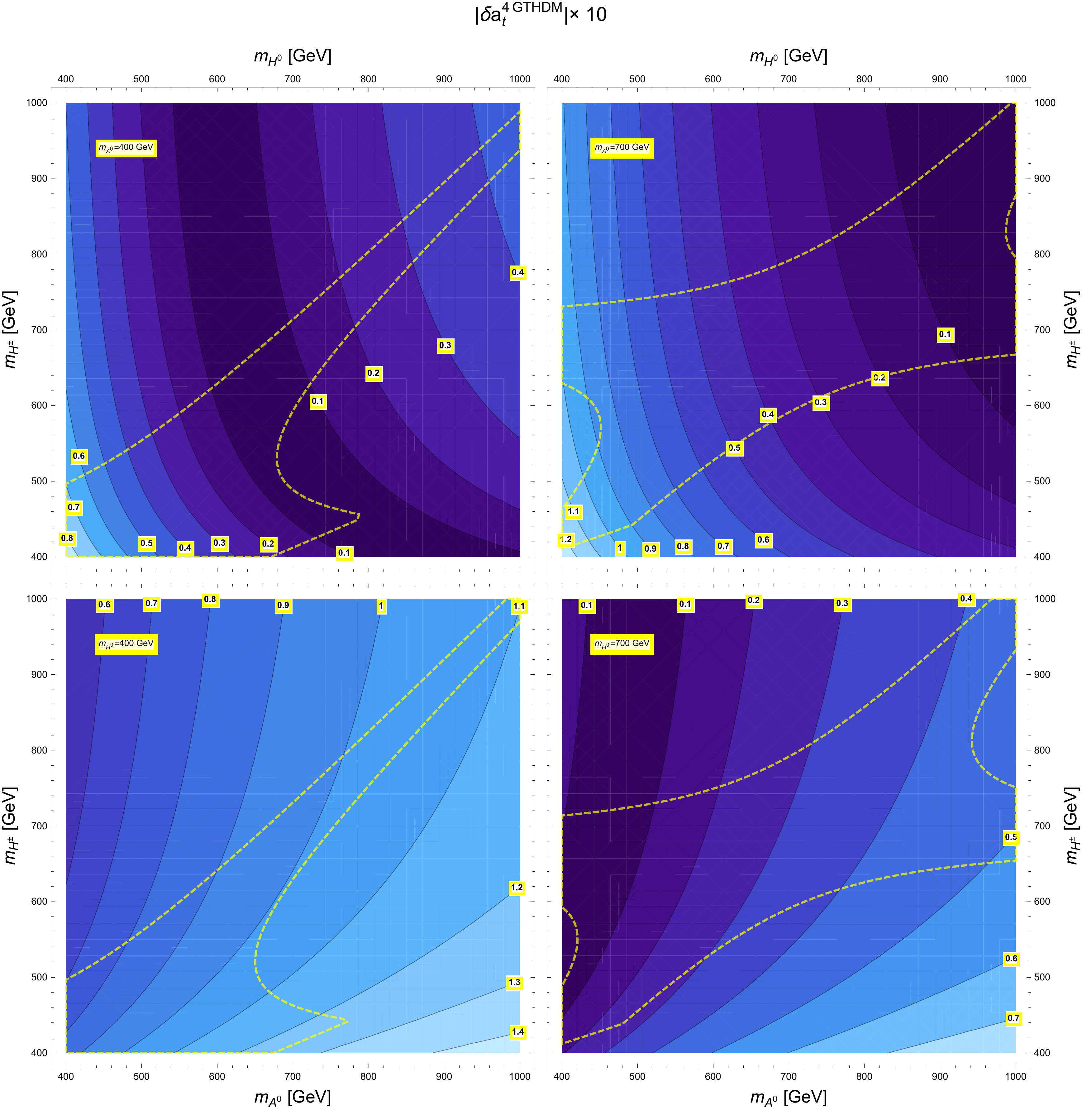}
  \caption{New physics contribution from the 4GTHDM to the top quark CMDM $a_t$ in units of $10^{-1}$. The top (bottom) plots show the $|a_t|$ contour lines in the $m_{H^0}$ vs $m_{H^\pm}$  ($m_{A^0}$ vs $m_{H^\pm}$) plane for the indicated  values of $m_{A^0}$ ($m_{H^0}$). We consider the parameter values of Table \ref{parameters} in the scenario with $(t_\beta,c_{\beta-\alpha})=(10,0.19)$. The area allowed by the constraints on the Higgs coupling modifiers  and the oblique parameters is enclosed by the dashed lines.}\label{ContourCMDM}
\end{figure}

To analyze the behavior of $a_t$ for nondegenerate scalar bosons, we  consider the scenario with $(t_\beta,c_{\beta-\alpha})=(10,0.19)$, which gives the largest $a_t$ values,  set $m_{b'}=350$ GeV, $\Delta_{t'-b'}=50$ GeV, and  use the values given in Table \ref{parameters} for the remaining  parameters of the model.  We then show in the top plots of Fig. \ref{ContourCMDM} the contour lines of  $\delta a_t^{\rm 4GTHDM}$ in the $m_{H^0}$ vs $m_{H^\pm}$ plane for $m_{A^0}=400$ and $700$ GeV, whereas in the bottom plots we show the corresponding contour  lines in the $m_{A^0}$ vs $m_{H^\pm}$ plane for  $m_{H^0}=400$ and $700$ GeV. The dashed lines enclose the  areas consistent with the constraints on the Higgs coupling modifiers and the oblique parameters. We observe in these  plots that there is a slight dependence of $a_t^{\rm 4GTHDM}$ on mild variations of the scalar boson masses, with the largest values of $\delta a_t^{\rm 4GTHDM}$  reached in three scenarios: relatively light degenerate scalar bosons ($m_{H^0}\sim  m_{H^\pm}\sim m_{A^0}\sim 400$ GeV); both $A^0$ and $H^\pm$  heavy and $H^0$ light ($m_{H^\pm}\sim m_{A^0}\sim 900$ GeV, $m_{H^0}\sim 400$ GeV); both $A^0$ and $H^\pm$  light and $H^0$ heavy  (  $m_{H^\pm}\sim m_{A^0}\sim 400$ GeV, $m_{H^0}\sim 900$ GeV ). There can also be an increase of $a_t^{\rm 4GTHDM}$ in other regions of the parameter space, which however are not compatible with the constraints on the Higgs coupling modifiers and the oblique parameters. In contrast, the smallest values of $a_t^{\rm 4GTHDM}$, of the order of $10^{-2}$, are reached in the regions where either all the three scalar bosons are heavy ($m_{H^\pm}\sim m_{A^0}\sim m_{H^0}> 700$ GeV) or $A^0$ is light and both $H^0$ and $H^{\pm}$ are heavy ($m_{A^0}\sim 400$ GeV $m_{H^0}\sim m_{H^\pm}\sim 700$ GeV).
 In general $\delta a_t^{\rm 4GTHDM}$ can be of the order of $10^{-2}$--$10^{-1}$, with a slight variation over the interval 400 GeV $\le m_\phi\le$ 1000 GeV.

We also have  analyzed the behavior of $\delta a^{\rm 4GTHMD}_t$  as a function of $m_{t'}$ for fixed  scalar boson masses  and the parameter values  given in Table \ref{parameters}. We have found that there is little dependence of  $\delta a^{\rm 4GTHMD}_t$ on $m_{t'}$ in the interval 350 GeV $\le m_{t'}\le$ 550 GeV, so we refrain from showing these results. This stems from the fact that the dominant contribution arises from $a_t^{\rm 3rd}$, which depends only on $m_{t'}$ through the charged Higgs boson contribution via the $H^- \bar{b}t$ coupling. However, this contribution is  smaller than those  of the heavy neutral Higgs bosons.
We also examined the dependence of $\delta a_t^{\rm 4GTHDM}$ on other   parameters of the model, but an enhancement above the $10^{-1}$ level was not found.

\subsubsection{Top quark CEDM}
\begin{figure}
  \centering
  \includegraphics[width=17cm]{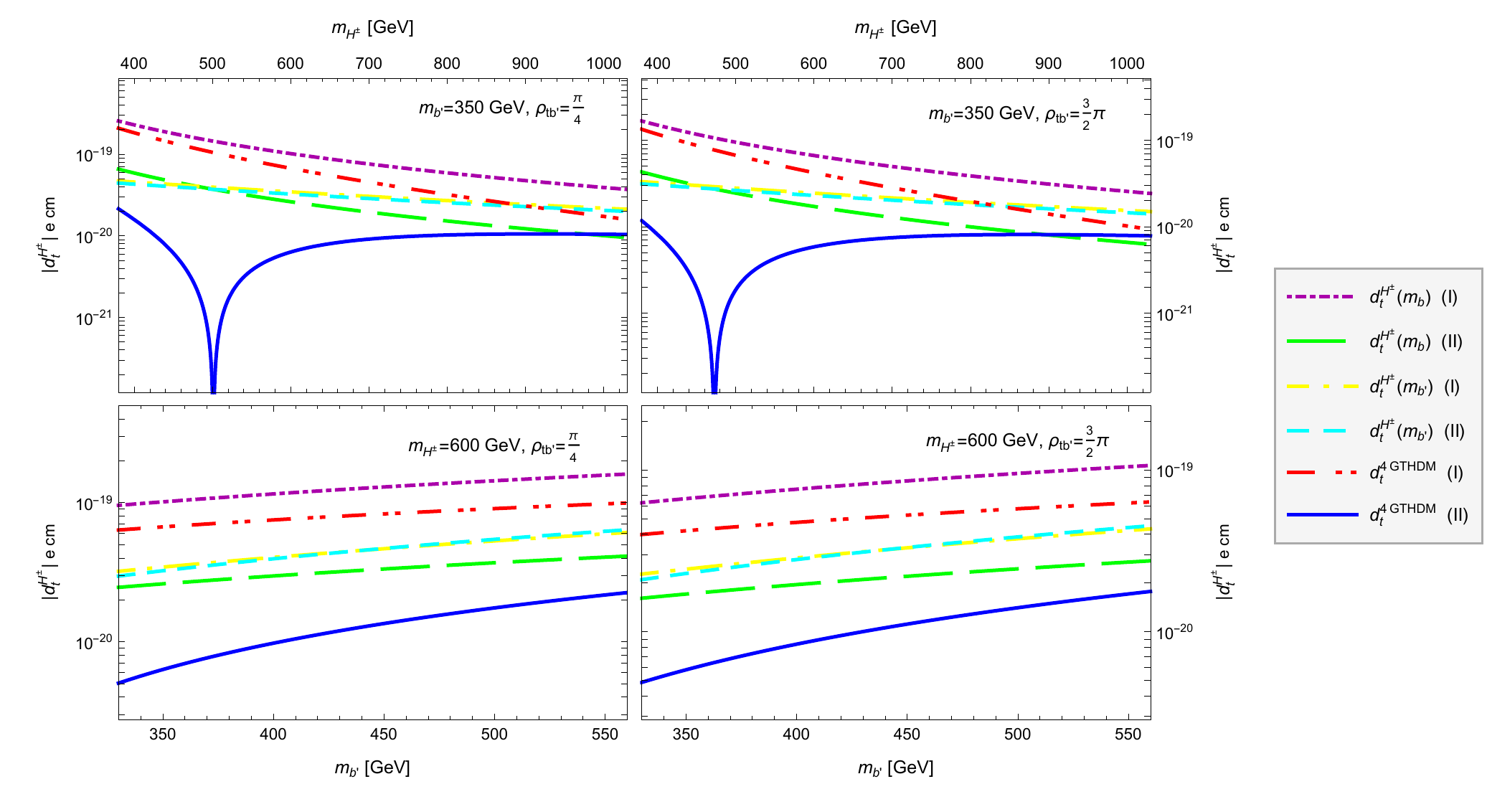}
  \caption{CEDM of the top quark in the 4GTHDM as a function of $m_{H^\pm}$ for fixed $m_{b'}$ (top plots) and as a function of $m_{b'}$ for fixed $m_{H^\pm}$ (bottom plots). We consider  two values of the complex phase $\rho_{tb'}$ and the  parameter values of Table \ref{parameters}, with $(t_\beta,c_{\beta-\alpha})=(10,0.19)$ (I) and $(t_\beta,c_{\beta-\alpha})=(5,0.4)$ (II). }\label{plotCEDM}
\end{figure}

As discussed above, new sources of $CP$  violation can arise in the 4GTHDM via the new phase of the extended CKM matrix but also through the mixing matrices $ \Sigma^{u,d}$. The analysis simplifies considerably since the contributions from the neutral scalar bosons to $d_t$ vanishes due to the hermicity of the mixing matrix $ \Sigma^{u,d}$, so  there is only contribution from the charged scalar boson. Therefore, $d_t$ is highly sensitive to the module and phase of the $
U_{tb'}$ and $U_{t'b}$ elements. The  4GTHDM contribution to the top quark CEDM can  be written as

\begin{equation}\label{CEDMTot}
d_t^{\rm 4GTHDM}=d_t^{H^\pm}(m_{b})+d_t^{H^\pm}(m_{b'}).
\end{equation}
In Fig. \ref{plotCEDM} we show the CEDM of the top quark in the 4GTHDM as a function of $m_{H^\pm}$   for fixed $m_{b'}$ (top plots) and as a function of  $m_{b'}$  for fixed $m_{H^\pm}$ (bottom plots). We consider  two values of the complex phase $\rho_{tb'}$ entering into the $4\times 4$ CKM mixing matrix and for the remaining parameters we use the values shown in Table \ref{parameters},  with  $(t_\beta,c_{\beta-\alpha})=(10,0.19)$ (scenario I) and $(t_\beta,c_{\beta-\alpha})=(5,0.4)$ (scenario II).
We first note that
in scenario I the dominant contribution to $d_t$ is that of the $b$ quark, with   the contribution of the $b'$ quark being slightly smaller. However, these partial contributions are of opposite signs and there are cancellation between them.
Therefore, in scenario I $d_t^{\rm 4GTHDM}$ can reach values of the order of $10^{-19}$ ecm for relatively light $m_{H^\pm}$, but it decreases up to $10^{-20}$ for $m_{H^\pm}=1$ TeV. As far as scenario II is concerned, both $b$ and $b'$ contributions are of similar size, but again  they can cancel each other out  (the large dip around $m_{H^\pm}=500$ GeV  in the top plots is due to the vanishing of $d_t$) so the total contribution can be rather suppressed, reaching values of the order of $10^{-20}$ ecm or below. In the bottom plots of Fig. \ref{plotCEDM} is evident that the largest contribution to $d_t^{\rm 4GTHDM}$ arises from the loop with the $b$ quark, which depends on $m_{b'}$ through the coupling $H^-\bar{b}t$. It is interesting to note that $d_t$ appears to increase as $m_{b'}$ increases. Along this line,  we have examined the behavior of $d_t$ [Eq. \eqref{CEDM}] for large $m_{H^\pm}$ and $m_{b'}$  in Appendix \ref{asymptotic} [see Eqs. \eqref{CMDMUHap} and \eqref{CEDMUHap}]. We have shown that $d_t$ decouples  as $m_{H^\pm}$ increases, but  there is nondecoupling  as $m_{b'}$ increases. However, our results cannot be considered valid for an ultra-heavy $b'$ quark as the 4GTHDM is a low-energy effective theory and unknown perturbative effects would give large corrections for $m_{b'}$ above 600 GeV.

\begin{figure}
  \centering
  \includegraphics[width=17cm]{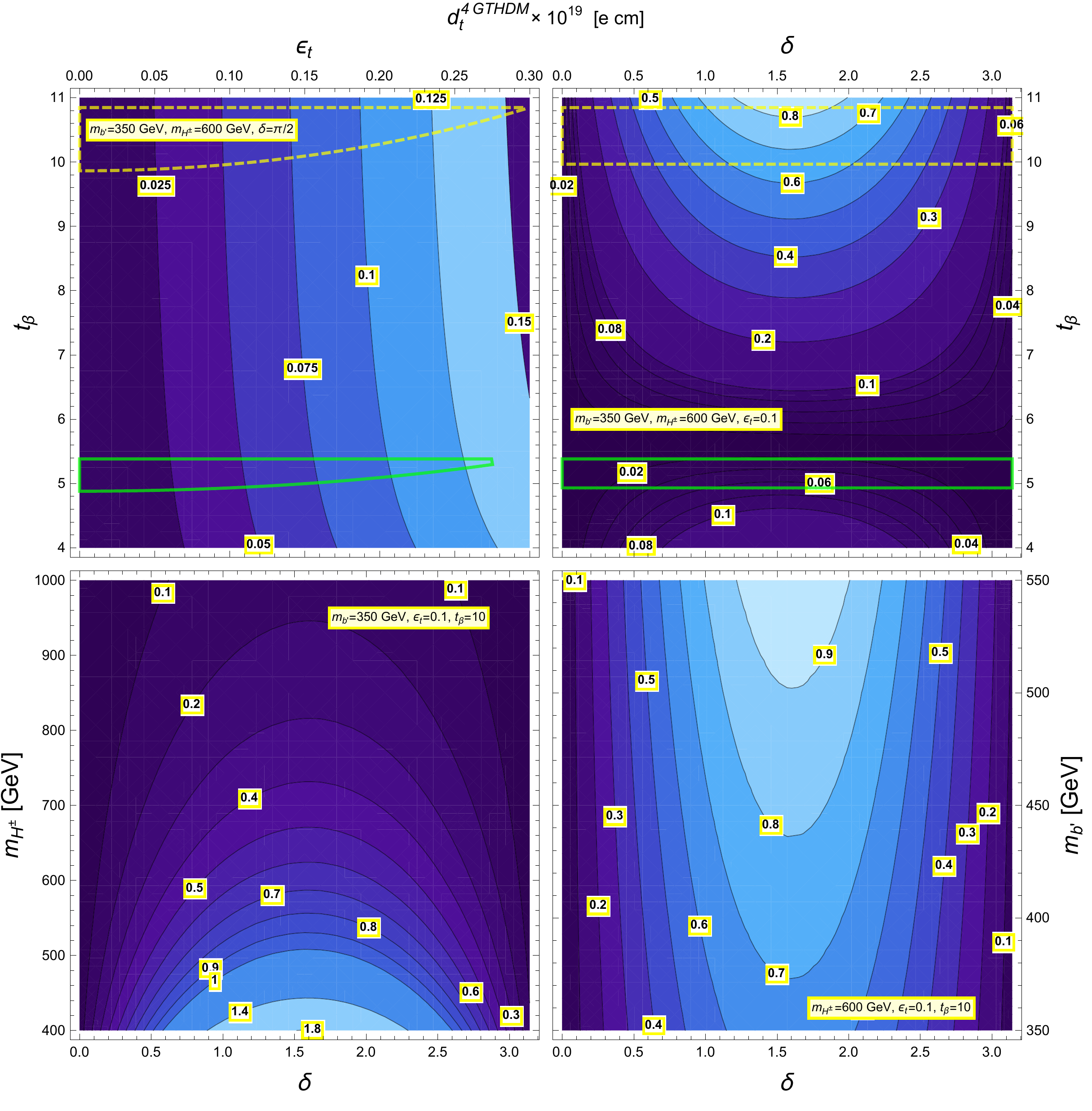}
  \caption{Contour lines of the  4GTHDM contribution to the top quark CEDM $d^{\rm 4GTHDM}_t$ in units of $10^{-19}$ ecm. We define $\delta=\rho_{tb'}+\delta_t$ and for the remaining parameters we consider the  values of Table \ref{parameters}, with $(t_\beta,c_{\beta-\alpha})=(10,0.19)$ in the bottom plots. The dashed and solid lines shown in the top plots enclose the area allowed by the LHC Higgs data when $c_{\beta-\alpha}=0.19$ and $c_{\beta-\alpha}=0.4$, respectively. }\label{ContourCEDM}
\end{figure}

We now analyze the dependence of $d_t$ on $t_\beta$ and the complex phases.
We found that there is little dependence on the  phase $\delta_b$ appearing in $\Sigma^d$, so  we refrain from presenting a detailed analysis along this line and focus instead on the dependence on $\rho_{tb'}$ and $\delta_t$, the complex phases of $U$ and  $ \Sigma^u$, respectively. Since  these phases can  interfere, we introduce the phase $\delta=\rho_{tb'}+\delta_t$. We first show in the top   plots of Fig. \ref{ContourCEDM} the contour lines of $d_t^{\rm 4GTHDM}$ in the $\epsilon_t$ vs $t_\beta$ and  $\delta$ vs $t_\beta$  planes, for the indicated parameter values. We also show the areas allowed by the LHC Higgs data for $c_{\beta-\alpha}=0.19$ (dashed line) and $c_{\beta-\alpha}=0.4$ (solid line). In the top-left plot we observe that for  a charged scalar boson with a mass $m_{H^\pm}=600$ GeV, $d_t^{\rm 4GTHDM}$ values of the order of $10^{-21}$  ecm ($10^{-20}$ ecm) can be reached for $|\epsilon_t|=0.1$ ($|\epsilon_t|=0.3$), with slightly larger values for $c_{\beta-\alpha}=0.19$. As for the top-right plot, we observe that $d_t^{\rm 4GTHDM}$ reaches its largest values for $\delta=\pi/2$ and large $t_\beta$, whereas  its smallest values are reached for $\delta=0,\,\pi$ and small $t_\beta$.

We now turn to  the bottom plots of Fig. \ref{ContourCEDM}, where we show the contour lines of $d_t^{\rm 4GTHDM}$ in the planes $\delta$ vs $m_{H^\pm}$ (bottom-left plot) and  $\delta$ vs $m_{b'}$ (bottom-right plot) for $(t_\beta,c_{\beta-\alpha})=(10,0.19)$ and the parameter values of Table \ref{parameters}.
We note  that $d_t$ reaches its largest values, of the order of $10^{-19}$ ecm,  for $\delta=\pi/2$ and either $m_{H^\pm}$ relatively light or   $m_{b'}$ close to its upper bound. We thus conclude that $d_t^{\rm 4GTHDM}$ can have values not much larger than about $10^{-19}$ ecm for $\delta=\pi/2$, $m_{H^\pm}$ relatively light, and  $m_{b'}$ close to its upper bound in the scenario with $(t_\beta,c_{\beta-\alpha})=(10,0.19)$, but values one order of magnitude smaller are reached in the scenario with $(t_\beta,c_{\beta-\alpha})=(5,0.4)$.  Also, there is little variation of $d_t$   with respect to other parameters such as $\epsilon_b$ and $\Delta_{t'-b'}$.

Finally, it is worth comparing the results for the CMDM and CEDM of the top quark in the 4GTHDM with the predictions of other popular extension models. In Table \ref{CMDMPred} we show the corresponding predictions, if available, of the top quark CMDM and CEDM in the usual THDMs, multiple Higgs-doublet models (MHDMs), 331 models, technicolor, extra dimensions, little Higgs models, SUSY theories, unparticles, and models with vectorlike multiplets. We conclude that the 4GTHDM can give contributions to the CMDM  of similar order of magnitude than these extension models, though the contribution to the CEDM can be larger than that predicted by the usual THDMs by one  order of magnitude, which is in part due to the presence of the new  quarks  via the corrections to the $H^-\bar{b}t$ coupling.

\begin{table}[htb!]
\begin{center}
\centering
\caption{Estimated order of magnitude of the CMDM and CEDM of the top quark in several extension models. MHDM stands for multiple Higgs doublet models with $CP$ violation.  The empty cells indicate that there is no known estimate in the corresponding model to our knowledge.}
\begin{tabular}{cccc}
\hline
\hline
 Model & $a_t$ && $d_t$ [ecm]   \\
\hline
\hline
SM & $10^{-2}$ \cite{ManyModels}&&   \\
\hline
THDMs & $10^{-3}$--$10^{-1}$ \cite{ManyModels,Gaitan:2015aia,THDMandMSSM} &&$10^{-20}$ \cite{Iltan:2001vg, Gaitan:2015aia}\\
\hline
{4GTHDM}&{$10^{-2}$--$10^{-1}$ }&&{$10^{-20}$--$10^{-19}$}
\\
\hline
MHDMs&&&$<10^{-19}$ \cite{Atwood:2000tu}\\
\hline
331 & $10^{-5}$ \cite{ManyModels} &&\\
\hline
Technicolor & $10^{-2}$ \cite{ManyModels}&& \\
\hline
Extra dimensions & $10^{-3}$ \cite{ManyModels}&& \\
\hline
Little Higgs & $10^{-6}$ \cite{LittleHiggs}&& \\
\hline
MSSM &$10^{-1}$ \cite{THDMandMSSM}  && $<10^{-19}-10^{-20}$ \cite{Atwood:2000tu} \\
\hline
Unparticles &$ 10^{-2}$ \cite{Unparticle}  &&  \\
\hline
 vectorlike Multiplets &  && $10^{-19}$ \cite{VecMul} \\
\hline
\hline
\end{tabular}
\label{CMDMPred}
\end{center}
\end{table}

\section{Conclusions and outlook}
\label{conclusions}
We have presented a calculation of the one-loop contributions to the chromomagnetic and chromoelectric dipole moments of the top quark within the two-Higgs doublet model with four fermion families, which predicts  new sources of $CP$ violation arising from the complex phases of two mixing matrices and the extended $4\times 4$ CKM matrix. Unlike the standard model with a sequential fourth generation of fermions, which is already excluded by the LHC data on the SM Higgs boson, there are still some regions of the parameter space of the 4GTHDM that are compatible with such data. These regions depend on the specific assumptions made for the parameters of the model. In particular we focus our calculation  on two regions still allowed by current constraints: the first one with $0.1  \lesssim t_\beta \lesssim 0.6$ and another one with $3 \lesssim t_\beta \lesssim 14$. The new contributions to the CMDM  of the top quark  arise from loops carrying the  new neutral scalar bosons   $H^0$ and $A^0$ accompanied by the  $t$ quark and the fourth-generation  $t'$ quark, together with loops carrying the charged scalar boson $H^\pm$ along with the $b$ quark and the fourth-generation $b'$ quark. There are also   new contributions from the lightest scalar boson $h^0$, which is identified with the SM Higgs boson, via   loops carrying the  $t$ and $t'$ quarks, with the former arising from the new physics correction to the $h\bar{t}t$ coupling. On the other hand, the CEDM of the top quark only receives the contribution from loops with the charged scalar boson along with the $b$ and $b'$ quarks. We present analytical expressions for all these contributions in terms of Feynman parameter integrals, which are explicitly integrated, and Passarino-Veltman scalar functions. We focus our numerical analysis of the behavior of the CMDM and CEDM of the top quark on the region of the parameter space of the 4GTHDM that is  still consistent with the LHC data on the 125 GeV Higgs boson and the experimental bounds on the oblique parameters. In particular we considered two scenarios for $t_\beta$ and $c_{\beta-\alpha}$: $(t_\beta,c_{\beta-\alpha})=(10,0.19)$ and $(t_\beta,c_{\beta-\alpha})=(5,0.4)$. In such  regions the top quark CMDM  can reach values of the order of $10^{ -2}-10^{-1}$, with the dominant contribution arising from the loops with the heavy scalar bosons accompanied by the top quark, whereas  the fourth-generation quarks give a smaller contribution. As for the   top quark CEDM, in the scenario with $(t_\beta,c_{\beta-\alpha})=(10,0.19)$  the dominant contribution, of the order of $10^{-19}$  ecm, arises from the loop with the $b$ quark, whereas the loop with the $b'$ quark gives a slightly smaller contribution. On the contrary, when $(t_\beta,c_{\beta-\alpha})=(5,0.4)$ the contributions to $d_t$ of both the $b$ and $b'$ quarks are of similar size. In both scenarios the $b$ and $b'$ contributions are of  opposite signs and tend to cancel each other out, with the strongest cancellation occurring in the $(t_\beta,c_{\beta-\alpha})=(5,0.4)$ scenario, in which case the corresponding contribution to $d_t$ can be smaller than $10^{-20}$ ecm. In general, the top quark CEDM can  reach values of the order of $10^{-20}$--$10^{-19}$  ecm for relatively light $m_{H^\pm}$ and  $m_{b'}$ heavy. Therefore, the  contributions arising from the 4GTHDM can be larger than those predicted by the usual THDM, which  is in part due to the presence of the new quarks via the corrections to the $H^-\bar{b}t$ coupling.

\acknowledgments{We acknowledge support from Consejo Nacional de Ciencia y Tecnolog\'ia and Sistema Nacional de Investigadores. Partial support from Vicerrector\'ia de Investigaci\'on y Estudios de Posgrado de la Ben\'emerita Universidad Aut\'onoma de Puebla is also acknowledged. }

\appendix

{
\section{Asymptotic behavior of the CMDM and CEDM   for heavy scalar bosons and heavy fourth-generation quarks}
\label{asymptotic}
We now examine the asymptotic behavior of the CMDM and CEDM of the top quark for ultra-heavy Higgs bosons and fourth-generation quarks in the 4GTHDM. After some algebra, Eqs. \eqref{CMDMintFP} and \eqref{CEDMintFP} can be integrated explicitly to give
\begin{align}
F(x,y)&=\frac{1}{\chi(x,y)} (y^2-(x+1)^2)  \left((1-2 x) x
   y^2+(x-1)^2 x (x+1)+y^4\right)
   (f(x,y)+f(y,x)\nonumber\\&+2 \left(x (2 x+1)
   y^2+(1-x) x (x+1)^2-y^4\right) \ln
   \left(\frac{x}{y}\right)+ 2 x (x+1)-2
   y^2-1,
\end{align}
and
\begin{align}
G(x,y)&=\frac{x}{\chi(x,y)}\left(x^4-2 x^2
   \left(y^2+1\right)+y^4+1\right)
   (f(x,y)+f(y,x)+
   2x\left(\left(x^2-y^2-1\right) \ln
   \left(\frac{x}{y}\right)-1\right),
\end{align}
where
\begin{align}
f(x,y)={\rm arctanh}\left(\frac{1-x^2+y^2}{\chi(x,y)
   }\right),
\end{align}
and
\begin{equation}
\chi(x,y)=\sqrt{\left((x-y)^2-1\right)
   \left((x+y)^2-1\right)}.
\end{equation}

For $x$ and $y$ very large we  can approximate the integrals  \eqref{CMDMintFP} and \eqref{CEDMintFP}  as
\begin{equation}
\label{Fapp}
F(x,y)=-G(x,y)\simeq \frac{x}{2
   \left(x^2-y^2\right)^3} \left(x^4-4 x^2 y^2+4 y^4
   \ln
   \left(\frac{x}{y}\right)+3
   y^4\right),
\end{equation}
which means that for $y\gg x$ (an ultra-heavy Higgs boson) we obtain

\begin{equation}
F(x,y)=-G(x,y)\simeq -\frac{4x \ln(y)}{2 y^2},
\end{equation}
whereas for  $x\gg y$ (an ultra-heavy fourth-generation quark) we have
\begin{equation}
F(x,y)=-G(x,y)\simeq \frac{1}{2x}.
\end{equation}
Using these expressions it is evident that both $a_t$ and $d_t$ behave as $1/m_{\phi}$ for large $m_\phi$ and thus decouple for an ultra-heavy internal Higgs boson.
For an ultra-heavy fourth-generation quark we need a more detailed analysis. We have for large $m_Q$ (neglecting $m_\phi$):
\begin{align}
a_{t}^{\phi}(m_Q) &\simeq\left(\frac{gm_t}{2 m_{W}} \right)^{2}\frac{|f_{\phi}|^{2} }{16\pi^{2}} \frac{\left( | {S}_{tQ}^\phi | ^2-| {P}_{tQ}^\phi | ^2\right)}{m_t\,m_Q}
,\label{CMDMUH}
\\
d_t^{H^\pm}(m_Q) &\simeq-\frac{g_s}{m_t}\left(\frac{gm_t}{2 m_{W}} \right)^{2}\frac{|f_{H^\pm}|^{2}} {16\pi^{2}}   \frac{\text{Im}\big({S}_{tQ}^{H^\pm} { P}_{tQ}^{H^\pm \ast} \big)}{m_t\,m_Q}\label{CEDMUH},
\end{align}
where $S^{\phi}_{ij}$ and $P^{\phi}_{ij}$ are given in Table \ref{Couplings}. For a neutral Higgs boson $h^0$, $H^0$, and $A^0$ and an internal ultra-heavy fourth-generation quark $Q=t'$,  we have $|{S}_{tt'}^{\phi}|\simeq |{P}_{tt'}^{\phi}|$. Therefore $a_t$ vanishes automatically and there is decoupling for large $m_{t'}$. Let us now examine  the contribution to $a_t$ and $d_t$ arising from the charged Higgs boson along with the $b'$ quark. According to the parametrization given in \eqref{Sigma_d} together with the definitions $\epsilon_t=\sin\theta_{tt'}e^{i\delta_t}$,  $U_{tb'}=|U_{tb'}|e^{i \rho_{tb'}} $, and   $U_{t'b}=|U_{tb'}|$, we obtain after some lengthy algebra

\begin{align}
| {S}_{tb'}^\phi | ^2-| {P}_{tb'}^\phi | ^2&\simeq -\frac{1}{s_\beta^2}
m_{b'}m_{t'}|U_{tb'}|| U_{t'b'}|\sin\theta_{tt'}\cos\delta,
\\
\text{Im}\big({S}_{tb'}^{H^\pm} { P}_{tb'}^{H^\pm \ast} \big)&\simeq\frac{1}{2}\frac{1}{s_\beta^2}
m_{b'}m_{t'}|U_{tb'} ||U_{t'b'}|\sin\theta_{tt'}\sin\delta,
\end{align}
with $\delta=\delta_t+\rho_{tb'}$. Thus $a_t$ and $d_t$ seem to behave as $m_{t'}$ for large $m_Q=m_{b'}$. However,  it is natural to assume that  $\sin\theta_{tt'}=O(m_t/m_{t'})$ \cite{1254G2HDM}, thus for large $m_{b'}$ we have
 \begin{align}
a_{t}^{H^\pm}(m_{b'}\gg m_t) &\simeq-\left(\frac{gm_t}{2 m_{W}} \right)^{2}\frac{1}{8\pi^{2}s_\beta^2}
|U_{tb'}|| U_{t'b'}|\cos\delta
,\label{CMDMUHap}
\\
d_t^{H^\pm}(m_{b'}\gg m_t) &\simeq-\frac{g_s}{m_t}\left(\frac{gm_t}{2 m_{W}} \right)^{2}\frac{1}{16\pi^{2}s_\beta^2}
|U_{tb'} ||U_{t'b'}|\sin\delta\label{CEDMUHap}.
\end{align}
We then conclude that there is nondecoupling for large $m_{b'}$ in the charged Higgs contribution.

Finally, we would like to note that \eqref{Fapp} can be written as

\begin{equation}\label{Fapp1}
F\left(\frac{m_Q}{m_t},\frac{m_\phi}{m_t}\right)\simeq
\frac{\sqrt{r s} \left((r-3)(r-1)+
   \ln (r)\right)}{2 (r-1)^3},
\end{equation}
with $r=m_Q^2/m_\phi^2\gg s=m_t^2/m_\phi^2$. Our approximate result agrees with previous results for the electron electric dipole moment, which involves a similar function
 \cite{Bernreuther:1990jx}.
}

\section{Oblique parameters $S$ and $T$ in  the 4GTHDM}
\label{oblique}
In the 4GTHDM the oblique parameters $S$ and $T$  receive new contributions from the fourth-generation fermions and the heavy scalar bosons \cite{Haber:1999zh,He:2001tp,Eberhardt:2010bm,Dighe:2012dz,Bar-Shalom:2016ehq}. They  can be written as

\begin{align}
S^{\rm 4GTHDM}&=S_F+S_\phi,
\end{align}
where $S_F$ stands for the contribution of the fourth generation of fermions and $S_\phi$ for the contribution of the heavy scalar bosons. Similar expressions are obeyed by $T^{\rm 4GTHDM}$. The corresponding expressions for the fourth-generation fermions are \cite{Eberhardt:2010bm,Dighe:2012dz}

\begin{equation}
S_F=\frac{1}{2\pi}\left(1-\frac{1}{6} \ln\left(\frac{m_{t'}^2}{m_{b'}^2}\right)\right)+\frac{1}{6\pi}\left(1+\frac{1}{2} \ln\left(\frac{m_{\nu'}^2}{m_{\ell'}^2}\right)\right),
\end{equation}
and
\begin{equation}
T_F=\frac{3}{8\pi s_W^2 m_W^2}\Bigg( |U_{t'b'}|^2F(m^2_{t'},m^2_{b'})+|U_{t'b}|^2F(m^2_{t'},m^2_{b})+|U_{tb'}|^2F(m^2_{t},m^2_{b'}])-|U_{tb}|^2F(m^2_t,m^2_b)+\frac{1}{3} F(m^2_{\ell'},m^2_{\nu'})\Bigg),
\end{equation}
where
\begin{equation}
F(x,y)=\left\{\begin{array}{lcr}
\frac{x+y}{2}-\frac{xy}{x-y}\ln\left(\frac{x}{y}\right)&&x\ne y,\\
0&&x=y.
\end{array}\right.
\end{equation}
As far as the heavy scalar bosons are concerned, they  give similar contributions to those arising in  the usual THDMs. The corresponding expressions in terms of Passarino-Veltman integrals are \cite{Haber:1999zh}
\begin{align}
S_{\phi}&=\frac{1}{\pi{m}_Z^2}\Bigg(s^2_{\beta-\alpha}
{\cal B}_{22}(m_Z^2, m^2_{H^0},m^2_{A^0})-
{\cal B}_{22}(m_Z^2, m^2_{H^{\pm}} ,m^2_{H^{\pm}} )+c^2_{\beta-\alpha}\left[{\cal B}_{22}(m^2_Z, m^2_{h^0},m^2_{A^0})+
{\cal B}_{22}(m^2_Z,m^2_Z,m^2_{H^0})\right.\nonumber\\
&\left.-{\cal B}_{22}(m^2_Z,m^2_Z,m^2_{h^0})
-m^2_Z{\cal B}_0(m^2_Z,m^2_Z,m^2_{H^0})+m^2_Z
{\cal B}_0(m^2_Z,m^2_Z,m^2_{h^0})\right]\Bigg),
\end{align}
and
\begin{align}
T_{\phi}&=\frac{1}{16\pi{m}^2_Ws^2_W}
\Bigg(F(m^2_{H^{\pm}} ,m^2_{A^0})+s^2_{\beta-\alpha}\left[
F(m^2_{H^{\pm}} ,m^2_{H^0})-F(m^2_{A^0},m^2_{H^0})\right]\nonumber\\
&+ c^2_{\beta-\alpha}\left[F(m^2_{H^{\pm}} ,m^2_{h^0})-F(m^2_{A^0},m^2_{h^0})
                             +F(m^2_W,m^2_{H^0})-F(m^2_W,m^2_{h^0})\right.
\nonumber\\
&   \left.
           -F(m^2_Z,m^2_{H^0})+F(m^2_Z,m^2_{h^0})
           +4m^2_Z \bar{B}_0(m^2_Z,m^2_{H^0},m^2_{h^0})
           -4m^2_W \bar{B}_0(m^2_W,m^2_{H^0},m^2_{h^0})   \right]\Bigg),
\end{align}
where
\begin{equation}
{\cal B}_{0}(q^2,m^2_1,m^2_2)
= B_{0}(q^2,m^2_1,m^2_2)-B_{0}(0,m^2_1,m^2_2),
\end{equation}
\begin{equation}
\bar{B}_0(m_1^2,m_2^2,m_3^2)
= B_{0}(0,m^2_1,m^2_2)-B_{0}(0,m^2_1,m^2_3),
\end{equation}
and
\begin{equation}
{\cal B}_{22}(q^2,m^2_1,m^2_2)
= B_{22}(q^2,m^2_1,m^2_2)-B_{22}(0,m^2_1,m^2_2).
\end{equation}
Analytical expressions for these Passarino-Veltman integrals were presented in \cite{He:2001tp} and are given as follows
\begin{equation}
{\cal B}_{0}(q^2,m^2_1,m^2_2)
=\left\{\begin{array}{lcr}1+\frac{1}{2}\left(
\frac{x_1+x_2}{x_1-x_2}-(x_1-x_2)\right)\ln\left(\frac{x_1}{x_2}\right)
+\frac{1}{2}f(x_1,x_2)&&x_1\ne x_2,\\
\\
 2 -2\sqrt{4x_1-1}\arctan\left(\frac{1}{\sqrt{4x_1-1}} \right)&&x_1=x_2,
 \end{array}\right.
 \end{equation}
where $x_i = m_i^2/q^2$, and
\begin{equation}
f(x_1,x_2)=\left\{
\begin{array}{lcr}
-2\sqrt{\Delta_{12}} \left(\arctan\left(\frac{x_1-x_2+1}{\sqrt{\Delta_{12}} }\right)
-\arctan\left(\frac{x_1-x_2-1}{\sqrt{\Delta_{12}} }\right)\right)&&\Delta_{12}>0,\\
0&&\Delta_{12}=0,\\
\sqrt{-\Delta_{12}} \ln\left(\frac{x_1+x_2-1+\sqrt{-\Delta_{12}} }{x_1+x_2-1-\sqrt{-\Delta_{12}} }\right)&&\Delta_{12}<0,
\end{array}
\right.
\end{equation}
with $\Delta_{12}=2(x_1+x_2)-(x_1-x_2)^2-1$. In addition
\begin{equation}
\bar{B}_0(m_1^2,m_2^2,m_3^2)=
\frac{m^2_1\ln{m^2_1}-m^2_3\ln{m^2_3}} {m^2_1-m^2_3}-
\frac{m^2_1\ln{m^2_1}-m^2_2\ln{m^2_2}} {m^2_1-m^2_2},
\end{equation}
\begin{align}
{\cal B}_{22}(q^2,m^2_1,m^2_2)&=\frac{q^2}{24}\Bigg(
2\ln{q^2}+\ln(x_1x_2)
+\left[(x_1-x_2)^3-3(x_1^2-x_2^2)
\right.\nonumber \\
&+\left.3(x_1-x_2)\right]\ln\left(\frac{x_1}{x_2}\right)
-\left[2(x_1-x_2)^2-8(x_1+x_2)+\frac{10}{3}\right]
\nonumber \\
&-\left[(x_1-x_2)^2-2(x_1+x_2)+1\right]f(x_1,x_2)
-6F(x_1,x_2)\Bigg),
\end{align}
for $m_1\ne m_2$, with
\begin{equation}
G(x)= -4\sqrt{4x-1} \arctan\frac{1}{\sqrt{4x-1}} ,.
\end{equation}
For  $m_2=m_1$, ${\cal B}_{22}(q^2,m^2_1,m^2_2)$ reduces to
\begin{equation}
{\cal B}_{22}(q^2,m^2_1,m^2_1)=\frac{q^2}{24}\left(2\ln{q^2} + 2\ln{x_1}
+\left(16x_1-\frac{10}{3}\right)+
\left(4x_1-1\right)G(x_1)\right).
\end{equation}

\bibliography{biblio}

\end{document}